\documentclass[fleqn,usenatbib]{mnras}

\usepackage{newtxtext,newtxmath}
\usepackage[T1]{fontenc}

\DeclareRobustCommand{\VAN}[3]{#2}
\let\VANthebibliography\thebibliography
\def\thebibliography{\DeclareRobustCommand{\VAN}[3]{##3}\VANthebibliography}

\usepackage{scalerel,tikz}
\usetikzlibrary{svg.path}
\definecolor{orcidlogocol}{HTML}{A6CE39}
\tikzset{orcidlogo/.pic={
 \fill[orcidlogocol] svg{M256,128c0,70.7-57.3,128-128,128C57.3,256,0,198.7,0,128C0,57.3,57.3,0,128,0C198.7,0,256,57.3,256,128z};
 \fill[white] svg{M86.3,186.2H70.9V79.1h15.4v48.4V186.2z}
 svg{M108.9,79.1h41.6c39.6,0,57,28.3,57,53.6c0,27.5-21.5,53.6-56.8,53.6h-41.8V79.1z M124.3,172.4h24.5c34.9,0,42.9-26.5,42.9-39.7c0-21.5-13.7-39.7-43.7-39.7h-23.7V172.4z}
 svg{M88.7,56.8c0,5.5-4.5,10.1-10.1,10.1c-5.6,0-10.1-4.6-10.1-10.1c0-5.6,4.5-10.1,10.1-10.1C84.2,46.7,88.7,51.3,88.7,56.8z};
}}
\newcommand\orcidicon[1]{\href{https://orcid.org/#1}{\mbox{\scalerel*{
\begin{tikzpicture}[yscale=-1,transform shape]
\pic{orcidlogo};
\end{tikzpicture}
}{|}}}}

\usepackage{graphicx}	% Including figure files
\usepackage{amsmath}	% Advanced maths commands
\usepackage{multirow}
\usepackage{array}
\usepackage{caption}

\newcommand{\vect}[1]{{\boldsymbol{#1}}}
\newcommand{\mach}{{\mathcal{M}}}
\newcommand{\cs}{{c_\mathrm{s}}}
\newcommand{\tturb}{t_\mathrm{turb}}
\newcommand{\eratio}{E_\mathrm{mag}/E_\mathrm{kin}}
\newcommand{\betak}{{\beta_\mathrm{k}}}

\title[Overcoming numerical dissipation at low Mach]{Mitigating numerical dissipation in simulations of subsonic turbulent flows}

\author[Watt et al.]{
James Watt,$^{1}$\thanks{E-mail: James.Watt@anu.edu.au}
Christoph Federrath$^{\orcidicon{0000-0002-0706-2306}}$,$^{1\thanks{E-mail: christoph.federrath@anu.edu.au}}$
Claudius Birke$^{2}$ and
Christian Klingenberg$^{2}$
\\
$^{1}$Research School of Astronomy and Astrophysics, Australian National University, Cotter Road, Canberra, ACT 2611, Australia\\
$^{2}$Department of Mathematics, University of W\"urzburg, W\"urzburg, Germany 
}

\date{Accepted XXX. Received YYY; in original form ZZZ}

\pubyear{2025}

\begin{document}
\label{firstpage}
\pagerange{\pageref{firstpage}--\pageref{lastpage}}
\maketitle

\begin{abstract}
    Magnetohydrodynamic (MHD) simulations of subsonic (Mach number~$<1$) turbulence are crucial to our understanding of several processes including  oceanic and atmospheric flows, the amplification of magnetic fields in the early universe, accretion discs, and stratified flows in stars. In this work, we demonstrate that conventional numerical schemes are excessively dissipative in this low-Mach regime. We demonstrate that a new numerical scheme (termed `USM-BK' and implemented in the FLASH MHD code) reduces the dissipation of kinetic and magnetic energy, constrains the divergence of magnetic field to zero close to machine precision, and resolves smaller-scale structure than other, more conventional schemes, and hence, is the most accurate for simulations of low-Mach turbulent flows among the schemes compared in this work. We first compare several numerical schemes/solvers, including Split-Roe, Split-Bouchut, USM-Roe, USM-HLLC, USM-HLLD, and the new USM-BK, on a simple vortex problem. We then compare the schemes/solvers in simulations of the turbulent dynamo and show that the choice of scheme affects the growth rate, saturation level, and viscous and resistive dissipation scale of the dynamo. We also measure the numerical kinematic Reynolds number (Re) and magnetic Reynolds number (Rm) of our otherwise ideal MHD flows, and show that the new USM-BK scheme provides the highest Re and comparable Rm amongst all the schemes compared.
\end{abstract}


%%%%%%%%%%%%%%%%%%%%%%%%%%%%%%%%%%%%%%%%%%%%%%%%%%

%%%%%%%%%%%%%%%%% BODY OF PAPER %%%%%%%%%%%%%%%%%%

\section{Introduction}

Subsonic flows are ubiquitous in a wide range of physical systems, ranging from terrestrial applications to astrophysics. They appear in the study of turbine blade performance \citep{LeggettZhaoSandberg2023}, fusion and fission systems \citep{MinEtAl2024}, rotorcraft fuselages and ship airwakes \citep{ParkLintonThornber2022}, ocean modelling \citep{TissotEtAl2023}, stratified systems like stars \citep{KupkaAndMuthsam2017}, and the amplification of primordial magnetic fields \citep{WagstaffEtAl2014, ChirakkaraEtAl2021}. Such subsonic flows are characterised by fluid velocities smaller than the speed of sound (also referred to as the low-Mach regime, where Mach number $M=v/c_\mathrm{s}<1$ and $v$ and $c_\mathrm{s}$ are flow velocity and sound speed, respectively). Being inherently non-linear and three-dimensional, these turbulent, complex systems are impossible to tackle via analytic calculations. Hence, they are studied through numerical simulations, a large class of which use finite volume discretisation and Godunov-based methods. While such methods are highly successful in modelling the transonic and the supersonic regime, they are subject to limitations in terms of efficiency in the subsonic regime. The artificial discontinuity created by the finite-volume method (FV method) at each cell interface creates spurious waves that lead to an overestimate of pressure, overwhelming the physical fluxes \citep[see][]{GuillardMurrone2004}, leading to excessive dissipation in the low-Mach regime. Apart from this, the discretisation of the MHD equations introduces viscous terms as well (see \citet[][]{ShivakumarAndFederrath2023}). Both of these effects combined operate similar to physical viscosity ($\nu$) and resistivity ($\eta$), and are referred to as artificial and numerical viscosity and resistivity. 

These types of numerical/artificial viscosity and resistivity must be significantly lower than the physical viscosity and resistivity simulations aiming to resolve down to the physical dissipation scale. Otherwise, small-scale features of the flow are smeared out. Flow properties in MHD are determined primarily by the hydrodynamic and the magnetic Reynolds numbers, labelled Re and Rm, respectively. They are defined as
\begin{equation}
    \mathrm{Re}=\frac{u\ell}{\nu},\, \text{and}
\end{equation}
\begin{equation}
    \mathrm{Rm}=\frac{u\ell}{\eta},
\end{equation}
where $u$ and $\ell$ are the characteristic velocity and length scales of the flow, respectively. To accurately model a flow, the numerical Re and Rm must be larger than the physical Re and Rm of the flows being modelled.

Numerical dissipation arising from discretisation can be reduced by increasing the grid resolution and special techniques, like Adaptive Mesh Refinement (AMR) \citep{BergerColella1989}. The effect of grid resolution on numerical viscosity and resistivity has been thoroughly studied by \citet{ShivakumarAndFederrath2023}. In this work, we focus on the dissipation originating from the artificial discontinuities created by the FV method in simulating subsonic flows. We test an extension of \citet{WagaanEtAl2011} introduced in \citet{BirkeKlingenberg2023}, which significantly reduces this dissipation.

In Section~\ref{sec:MHDEquations}, we introduce the MHD equations. Section~\ref{sec:methods} describes our numerical methods and briefly summarises why simulations of subsonic flows are more dissipative and how \citet{BirkeKlingenberg2023} overcome this difficulty. Section~\ref{sec:balsara} presents simulations of the Balsara vortex (\cite{Balsara2004}, see also \cite{LeidiEtAl2022}) as a test case for a variety of numerical schemes, all at the same grid resolution, to demonstrate the effect of the choice of numerical scheme on artificial viscosity and resistivity and test if the new scheme provides an improvement over previous methods. In Section~\ref{sec:Turbulent dynamo}, we test the various numerical schemes on simulations of subsonic turbulent dynamos, which also have important astrophysical applications. We compare the effect of the numerical scheme on the time evolution and saturation of the dynamo, as well as the structure and turbulent MHD statistics of the system. We also measure the numerical Reynolds numbers of these otherwise ideal MHD ($\mathrm{Re}\rightarrow\infty$, $\mathrm{Rm}\rightarrow\infty$) simulations. We summarise our results in Section~\ref{sec:Conclusion}.

% ==== Methods ====
\section{Equations of MHD}
\label{sec:MHDEquations}
The MHD equations are given as follows:
\begin{equation}
    \frac{\partial \rho}{\partial t}+\nabla\cdot(\rho\vect{u})=0,\label{eq:Compressibility}
\end{equation}
\begin{equation}
    \frac{\partial}{\partial t}\left(\rho\vect{u}\right)+\nabla\cdot\left(\rho\vect{u}{\otimes}\vect{u}-\frac{1}{4\pi}\vect{B}{\otimes}\vect{B}\right)+\nabla p_{\mathrm{{tot}}}=\nabla\cdot(2\nu\rho \vect{S})+\rho\vect{F},
\end{equation}
\begin{multline}
    \frac{\partial e}{\partial t}+\nabla\cdot\left[\left(e+p_{\mathrm{tot}}\right)\vect{u}-{\frac{1}{4\pi}}\left(\vect{B}\cdot\vect{u}\right)\vect{B}\right]=\\\nabla\cdot\left[2\nu\rho\vect{u}\cdot{\boldsymbol{S}}+{\frac{1}{4\pi}}\eta\vect{B}\times(\nabla\times\vect{B})\right],
\end{multline}
\begin{equation}
    \frac{\partial}{\partial t}{\vect{B}}=\nabla\times({\vect{u}}\times{\vect{B}})+\eta\nabla^{2}{\vect{B}},\\
\end{equation}
Here, $\rho$, $\vect{u}$, $p_\mathrm{tot}=p_\mathrm{th}+|\vect{B}|^2/(8\pi)$, $\vect{B}$, $e=\rho e_\mathrm{int} + \rho|\vect{u}|^2/2+|\vect{B}|^2/(8\pi)$, $S_\mathrm{ij}=(\partial_\mathrm{i}u_\mathrm{j}+\partial_\mathrm{j}u_\mathrm{i})/2-(\delta_\mathrm{ij}\nabla\cdot\vect{u})/3$, $\nu$, $\eta$ denote the density, velocity, pressure (thermal plus magnetic), magnetic field, energy density (internal plus kinetic, plus magnetic), strain tensor, kinematic viscosity and magnetic resistivity, respectively. The system of equations is closed by an equation of state relating the thermal pressure ($p_\mathrm{th}$) to the density ($\rho$).

The magnetic field also satisfies the divergence free constraint given by
\begin{equation}
    \nabla\cdot{\vect{B}}=0.\label{eq:DivB}
\end{equation}

In the absence of physical viscosity and resistivity ($\nu=0$ and $\eta=0$), the system of partial differential equations described above takes the general conservative form
\begin{equation}
    \frac{\partial \vect{U}}{\partial t} + \frac{\partial \vect{F}}{\partial x} + \frac{\partial \vect{G}}{\partial y} + \frac{\partial \vect{H}}{\partial z}=0,
    \label{eq:Conservative PDE}
\end{equation}
where $\vect{U}$ is a vector of conservative variables given by
\begin{equation}
    \vect{U} = \left[\rho,\, \rho u_\mathrm{x},\, \rho u_\mathrm{y},\, \rho u_\mathrm{z},\, e,\, B_\mathrm{x},\, B_\mathrm{y},\, B_\mathrm{z}\right],
\end{equation}
and $\vect{F}$, $\vect{G}$ and $\vect{H}$ are the fluxes given by 
\begin{multline}
    \vect{F} = [\rho v_{\mathrm{x}} ,\, \rho v_{\mathrm{x}}^2 + p_\mathrm{th} + |\vect{B}|^2 / 8 \pi - B_{\mathrm{x}}^2 / 4 \pi ,\, \rho v_{\mathrm{x}} v_{\mathrm{y}} - B_{\mathrm{x}} B_{\mathrm{y}} / 4 \pi ,\,\\ \rho v_{\mathrm{x}} v_{\mathrm{z}} - B_{\mathrm{x}} B_{\mathrm{z}} / 4 \pi ,\, \left( e + p_\mathrm{th} + |\vect{B}|^2 / 8 \pi \right) v_{\mathrm{x}} - B_{\mathrm{x}} (\mathbf{v} \cdot \mathbf{B}) / 4 \pi ,\, 0 ,\,\\ (v_{\mathrm{x}} B_{\mathrm{y}} - v_{\mathrm{y}} B_{\mathrm{x}}) ,\, -(v_{\mathrm{z}} B_{\mathrm{x}} - v_{\mathrm{x}} B_{\mathrm{z}})],
\end{multline}
\begin{multline}
    \vect{G} = [\rho v_{\mathrm{y}} ,\, \rho v_{\mathrm{x}} v_{\mathrm{y}} - B_{\mathrm{x}} B_{\mathrm{y}} / 4 \pi ,\, \rho v_{\mathrm{y}}^2 + p_\mathrm{th} + |\vect{B}|^2 / 8 \pi - B_{\mathrm{y}}^2 / 4 \pi ,\,\\
    \rho v_{\mathrm{y}} v_{\mathrm{z}} - B_{\mathrm{y}} B_{\mathrm{z}} / 4 \pi ,\, \left( e + p_\mathrm{th} + |\vect{B}|^2 / 8 \pi \right) v_{\mathrm{y}} - B_{\mathrm{y}} (\mathbf{v} \cdot \mathbf{B}) / 4 \pi ,\, 0 ,\,\\ 
    (v_{\mathrm{y}} B_{\mathrm{z}} - v_{\mathrm{z}} B_{\mathrm{y}}) ,\, -(v_{\mathrm{x}} B_{\mathrm{y}} - v_{\mathrm{y}} B_{\mathrm{x}})],\, \text{and}
\end{multline}
\begin{multline}
    \vect{H} = [\rho v_{\mathrm{z}} ,\, \rho v_{\mathrm{x}} v_{\mathrm{z}} - B_{\mathrm{x}} B_{\mathrm{z}} / 4 \pi ,\, \rho v_{\mathrm{y}} v_{\mathrm{z}} - B_{\mathrm{y}} B_{\mathrm{z}} / 4 \pi ,\,\\ 
    \rho v_{\mathrm{z}}^2 + p_\mathrm{th} + |\vect{B}|^2 / 8 \pi - B_{\mathrm{z}}^2 / 4 \pi ,\, \left( e + p_\mathrm{th} + |\vect{B}|^2 / 8 \pi \right) v_{\mathrm{z}} - B_{\mathrm{z}} (\mathbf{v} \cdot \mathbf{B}) / 4 \pi ,\\ 0 ,\,
    (v_{\mathrm{x}} B_{\mathrm{z}} - v_{\mathrm{z}} B_{\mathrm{x}}) ,\, -(v_{\mathrm{y}} B_{\mathrm{z}} - v_{\mathrm{z}} B_{\mathrm{y}})].
\end{multline}

\section{Numerical methods}
\label{sec:methods}

\subsection{Finite volume method}

Eq.~(\ref{eq:Conservative PDE}) can be solved using the finite-volume (FV) method. The FV method divides the computational domain into control volumes (grid cells) and integrates the governing equations over each volume, ensuring conservation of fluxes across cell boundaries. Fluxes at the interfaces are computed using Riemann solvers (such as Roe, HLLD, HLLC, etc). To improve accuracy, the physical state variables are reconstructed at the cell faces via linear or even higher-order reconstruction. Slope limiters are used to ensure that the reconstruction step does not introduce artificial maxima/minima. The temporal discretisation is performed using schemes such as Euler or Runge-Kutta methods and the time-stepping can be implemented in a split or unsplit fashion. The divergence of the magnetic field is constrained to zero up to machine precision using constrained transport (CT) \citep{Yee1966, EvansAndHawley1988, DaiAndWoodward1998, GardinerAndStone2008}, or kept at reasonably low levels by a divergence cleaning technique \citep{Dedner2002}.

\subsection{Numerical dissipation}

Discretisation of MHD equations gives rise to viscous terms that introduce numerical dissipation. This kind of numerical dissipation can be reduced by choosing a reconstruction method or a time stepper of higher order. However, finite volume methods also create an artificial discontinuity at each grid interface that gives rise to spurious waves. Since these artificial waves create pressure fluctuations of the order of the sonic Mach number $M$, they can overwhelm the physical flux in the simulations of low-Mach flows that have pressure fluctuations of the order $M^2$ \citep{GuillardMurrone2004}. This leads to excessive dissipation in low-Mach flows. Various methods have been explored to mitigate this issue, like pre-conditioning the Riemann problem at each interface to reduce the effect of discontinuities \citep{Turkel1999, Clerc2000}, or rescaling the dissipation term in the numerical flux to make it independent of the Mach number \citep{MiczekEtAl2015, MinoshimaAndMiyoshi2021, LeidiEtAl2022, BirkeChalonsKlingenberg2023}. Another approach is the use of implicit-explicit methods, which apply the Godunov-type scheme only to the slow dynamics in the PDE, thereby avoiding dissipation terms that scale with $O(1/M)$ \citep{Klein1995, Dumbser2018, Fambri2021, ChenEtAl2023, BirkeBoscheriKlingenberg2023, BoscheriThomann2024}. Building on this idea, \citet{FambriSonnendrucker2024} employ implicit-explicit methods in combination with the Finite Element method for solving the magneto-acoustic parts, ensuring energy stability, magnetic-helicity conservation, and a divergence-free magnetic field. Alternatively, \citet{TeissierMuller2024} reduce artificial and numerical viscosity through very high-order reconstruction methods, and improve efficiency by reconstructing separately in each spatial dimension rather than using multidimensional polynomials. In this work, we focus on the relaxation scheme by \citet{BirkeKlingenberg2023} (referred to as the BK method), which resorts on rescaling the numerical flux in the low-Mach-number regime.

\subsection{BK method}
\label{sec:BK method}

The core idea of the BK method is to construct an enlarged system of equations, including a relaxation term on the right-hand side, such that the new system is an approximation of the original system given in Section~\ref{sec:MHDEquations}. Then the left-hand side of the relaxation system is solved using a Riemann solver followed by a projection of the solution back onto the original variables. Since there is some freedom in how the relaxation system is constructed, it is possible to tweak the solution of the pressure variable in the Riemann fan and fix the incorrect scaling of the pressure, while ensuring that the resulting Riemann solver satisfies a discrete entropy inequality. We point the reader to \citet{BirkeKlingenberg2023} for further details on their relaxation scheme.

The fastest wave-speed in the Riemann fan of the BK relaxation scheme can be closely approximated by
\begin{equation} \label{eq:lambda_fastest}
    \lambda_\mathrm{fastest} = u+ \frac{1}{2}\sqrt{\left(\frac{c_\mathrm{s}^2}{M_\mathrm{BK}^2}+c_\mathrm{A}^2\right) +
    \sqrt{\left(\frac{c_\mathrm{s}^2}{M_\mathrm{BK}^2}+c_\mathrm{A}^2\right)^2-4c_\mathrm{s}^2c_\mathrm{A;x}^2}},
\end{equation}
where $u$ is the fluid velocity, $c_\mathrm{s}$ is the sound speed, $c_\mathrm{A}$ is the Alfv\'en speed, $c_\mathrm{A;x}$ is the Alfv\'en speed in the x-direction (direction along which the MHD equations are one-dimensionalised before solving the Riemann problem) and $M_\mathrm{BK}$ \citep[the equivalent of $\phi$ in Eq.~(14)--(16) in][]{BirkeKlingenberg2023} is defined as
\begin{equation} \label{eq:Mcut}
    M_\mathrm{BK} = \mathrm{min}\left\{\mathrm{max}\left\{M_\mathrm{cut}, \frac{u}{c_\mathrm{s}}\right\}, 1\right\}.
\end{equation}
The parameter $M_\mathrm{cut}<1$ is used to set a local cut-off Mach number below which the scheme does not reduce dissipation any longer by increasing the scheme-specific speed, thereby preventing division by small numbers in regions where the velocity is close to 0, which would lead to $\lambda_\mathrm{fastest}\to\infty$ and the time-step $\Delta t\to0$. In this work, we set $M_\mathrm{cut}$ equal to $2$ times the reference Mach number ($\mathcal{M}=\,$0.1 or 0.01) that we are simulating -- the reference Mach number is a statistical (global) quantity describing the characteristic ratio of typical flow velocities to the sound speed in a given problem\footnote{For instance, in turbulent flows, $\mathcal{M}$ is the standard deviation of the local Mach number ($M$).}. The reason for our choice of $M_\mathrm{cut}=2\,\mathcal{M}$ is explained in Appendix~\ref{appendix:Mcut}. Note that for a conventional Riemann solver, like Roe or HLLD, $M_\mathrm{BK}=1,$ and $\lambda_\mathrm{fastest}$ is a close approximation of the fastest wave speed in the Riemann fan of these conventional solvers.

The time-step restriction for stability is given by the Courant–Friedrichs–Lewy (CFL) condition,
\begin{equation}
    \Delta t = \mathrm{CFL}\frac{\Delta x}{\lambda_\mathrm{fastest}},
    \label{eq:CFL condition}
\end{equation}
where $\Delta x$ is the cell size. We use $\mathrm{CFL}=0.5$ throughout this work.

\subsection{Numerical schemes in FLASH}

\begin{table*}
    \centering
    \caption{Numerical schemes used in this work.}
    \label{tab:solvers}
    \begin{tabular}{lcccc} % four columns, alignment for each
        \hline
        Scheme Name & Split / Unsplit & Riemann Solver & $\nabla\cdot\vect{B}$ Method & Electric Field Reconstruction \\
        (1)           & (2)                      & (3)                     & (4)                                   & (5)                                    \\
        \hline
        Split-Roe     & Split                    & Roe                     & Dedner-Marder cleaning                & N/A                                    \\
        Split-Bouchut & Split                    & Bouchut                 & Dedner-Marder cleaning                & N/A                                    \\
        USM-Roe       & Unsplit                  & Roe                     & Constrained Transport                 & Lee-Upwind                             \\
        USM-HLLD      & Unsplit                  & HLLD                    & Constrained Transport                 & Lee-Upwind                             \\
        USM-HLLC      & Unsplit                  & HLLC                    & Constrained Transport                 & Lee-Upwind                             \\
        USM-BK        & Unsplit                  & BK                      & Constrained Transport                 & Lee-Upwind                             \\
        \hline
    \end{tabular}
    \begin{flushleft}
        \textbf{Notes.} Column~(1): name of numerical scheme, (2): whether the scheme uses directionally split or unsplit updates, (3): Riemann solver -- Roe \citep{Roe1981}, Bouchut \citep{WagaanEtAl2011}, HLLD \citep{MiyoshiKusano2005}, HLLC \citep{Li2005} or BK \citep{BirkeKlingenberg2023}, (4): whether Dedner-Marder cleaning \citep{Marder1987, Dedner2002} or constrained transport (CT) \citep{Yee1966, EvansAndHawley1988, DaiAndWoodward1998, GardinerAndStone2008} was used to handle the magnetic field divergence constraint, (5): Lee-upwind \citep{Lee2006} electric field reconstruction method, if applicable. All schemes use the 2nd-order Hancock \citep{Leer1984} method for interpolation of data in space and time.
    \end{flushleft}
\end{table*}

We perform our simulations using a modified version of the FLASH code \citep{Fryxell2000, DubeyEtAl2008} and compare several numerical schemes with different Riemann solvers. Some schemes utilise split time-stepping combined with Dedner-Marder cleaning \citep{Marder1987, Dedner2002} for magnetic field divergence control (Split-Roe and Split-Bouchut), while others adopt unsplit time-stepping on a staggered mesh with an upwind version of Lee's constrained transport scheme \citep{Lee2006} (USM-Roe, USM-HLLD, USM-HLLC, and USM-BK, where `USM' stands for `unsplit-staggered mesh'). All our schemes use the 2nd-order TVD (total variation diminishing) interpolation of data in space and time using the Hancock method \citep{Leer1984}. The details of the numerical schemes are summarised in Table~\ref{tab:solvers}.

\section{Balsara vortex} \label{sec:balsara}

The Balsara vortex \citep{Balsara2004} is an exact stationary solution of the ideal MHD equations in two dimensions, where the centrifugal force, magnetic tension, thermal pressure gradient, and magnetic pressure gradient are perfectly balanced. This configuration, which conserves kinetic and magnetic energies independently in the absence of dissipative forces, serves as an excellent test problem for evaluating energy conservation in MHD simulations. Discretisation errors and artificial discontinuities in finite-volume methods introduce numerical dissipation, leading to a loss of rotational and magnetic energy. Here we use the Balsara vortex to compare the energy conservation performance of various split and unsplit MHD solvers across different numerical schemes. While the dissipation we observe arises from a combination of numerical discretisation and the choice of numerical technique (like reconstruction method, Riemann solver, etc.) as well as the presence of artificial discontinuities, we use the same resolution for all the different schemes in Tab.~\ref{tab:solvers}, such that we can compare the dissipation arising from the latter.

\subsection{Initial conditions}

\begin{figure}
    \includegraphics[width=1.0\linewidth]{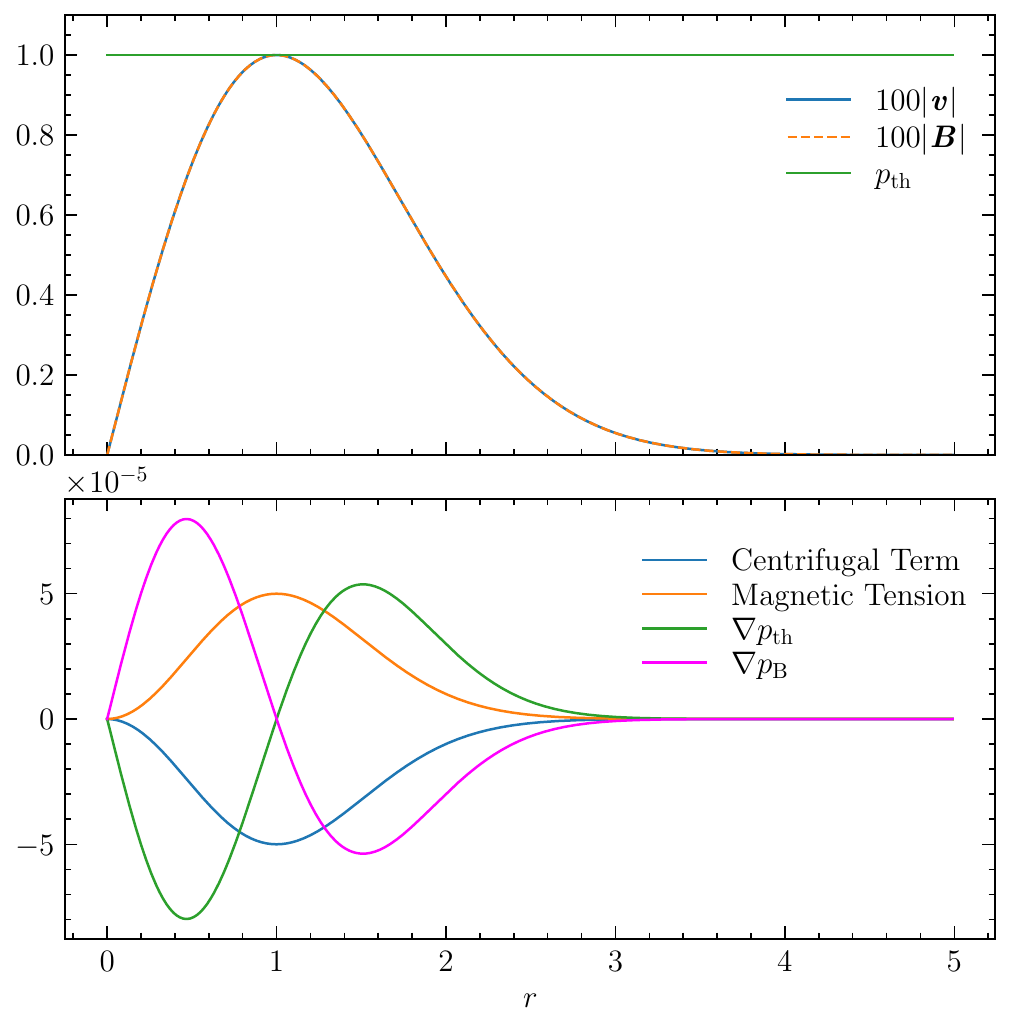}
    \caption{Top panel shows the radial profiles of velocity, magnetic field and pressure for the Balsara vortex, following Eqs.~(\ref{eq:BV_vel})--(\ref{eq:BV_pres}) for a sonic Mach number of $\mach=0.01$ and the ratio of the magnetic to the rotational kinetic energy $\betak=1$. Note that the velocity and magnetic pressure profiles have been scaled by a factor of 100 for the sake of clarity. The scaled velocity profile touches the thermal pressure profile ($p_\mathrm{th}\approx1$) at $r=1$ since $\mathcal{M}=0.01$. The bottom panel shows that the centrifugal term $-(\vect{v}\cdot\nabla)\vect{v}$ is balanced by the magnetic tension $(\vect{B}\cdot\nabla)\vect{B}$, and the gradients of the thermal pressure ($\nabla p_\mathrm{th}$) and the magnetic pressure ($\nabla p_\mathrm{B}$) balance each other.}
    \label{fig:initial_profile}
\end{figure}

The initial conditions for the Balsara vortex are given by
\begin{eqnarray}
    \vect{v} &=& \tilde{v}\left(-y\hat{\vect{x}}+x\hat{\vect{y}}\right)\exp\left(\frac{1-r^2}{2}\right), \label{eq:BV_vel}\\
    \vect{B} &=& \tilde{B}\left(-y\hat{\vect{x}}+x\hat{\vect{y}}\right)\exp\left(\frac{1-r^2}{2}\right), \\
    p_\mathrm{th} &=& 1+\left[\frac{\tilde{B}^2}{2}(1-r^2)-\frac{\tilde{v}^2}{2}\right]\exp\left({1-r^2}\right), \label{eq:BV_pres}\\
    \rho &=& 1, \label{eq:BV_rho}
\end{eqnarray}
where $r^2=x^2+y^2$, and $\hat{\vect{x}}$ and $\hat{\vect{y}}$ are unit vectors in the $x$ and $y$ directions, respectively. We use $\tilde{v}=0.01$, $\tilde{B}=0.01$, and $\gamma=5/3$. Here we define the reference Mach number ($\mathcal{M}$) as the maximum local Mach number ($M$) in the simulation domain. Our choice of parameters $\tilde{v}\,,\tilde{B}\,\text{and}\,\gamma$ gives $\mathcal{M}\approx0.01$. The radial profiles of velocity, magnetic field and pressure are shown in Fig.~\ref{fig:initial_profile}.

\subsection{Setup}

We use a computational domain of $(x,y)\in[-5,5]\times[-5,5]$ and $64\times64$ grid cells with periodic boundary conditions and $M_\mathrm{cut}=0.02$ for our simulations. The problem is made computationally harder by advecting the vortex along the diagonal of the computational grid with speed $\tilde{v}$. We run our simulations for one complete advection of the vortex across the diagonal, such that it ends up exactly at the starting position, i.e., at the coordinate origin. In this time interval, the vortex turns around $2.25$ times.

\subsection{Results and comparison of solvers}
\label{sec:BV results}

\begin{figure}
    \centering
    \includegraphics[width=1\linewidth]{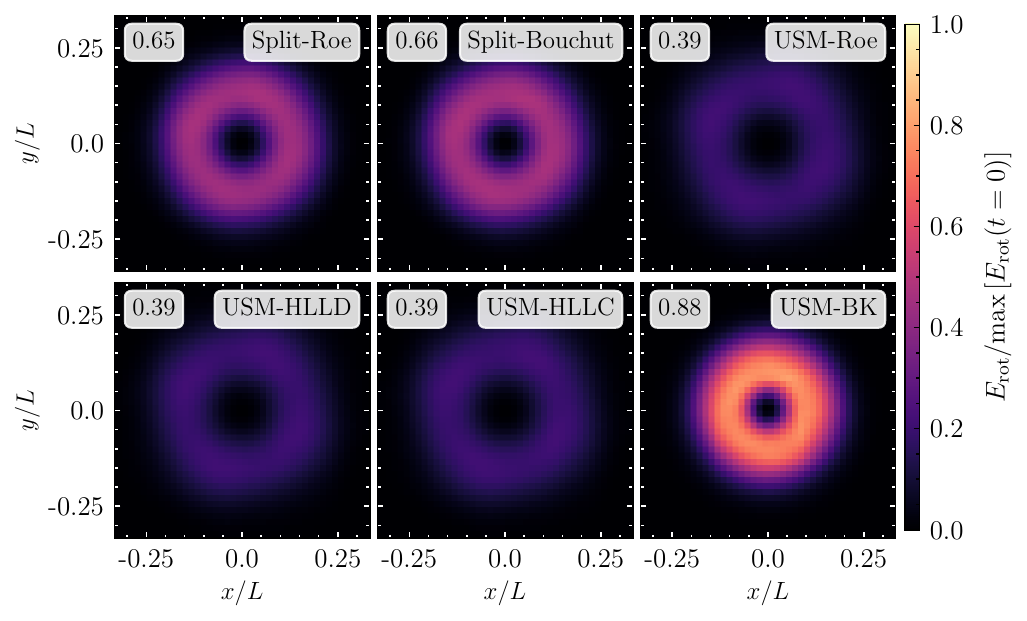}
    \caption{Rotational energy of the vortex after one advection diagonally through the computational domain for the six different numerical solver/scheme combinations (from left to right and top to bottom): Split-Roe, Split-Bouchut, USM-Roe, USM-HLLD, USM-HLLC, and USM-BK. The rotational energy has been normalised by the maximum rotational energy (at $r=1$) at the beginning of the simulation ($t=0$). The value in the top left corner of each panel shows the fraction of the total rotational energy left in the system compared to $t=0$. We see that USM-BK outperforms all other schemes by retaining 88\% of the rotational kinetic energy.}
    \label{fig:LowMachERot}
\end{figure}

\begin{figure}
    \includegraphics[width=1\linewidth]{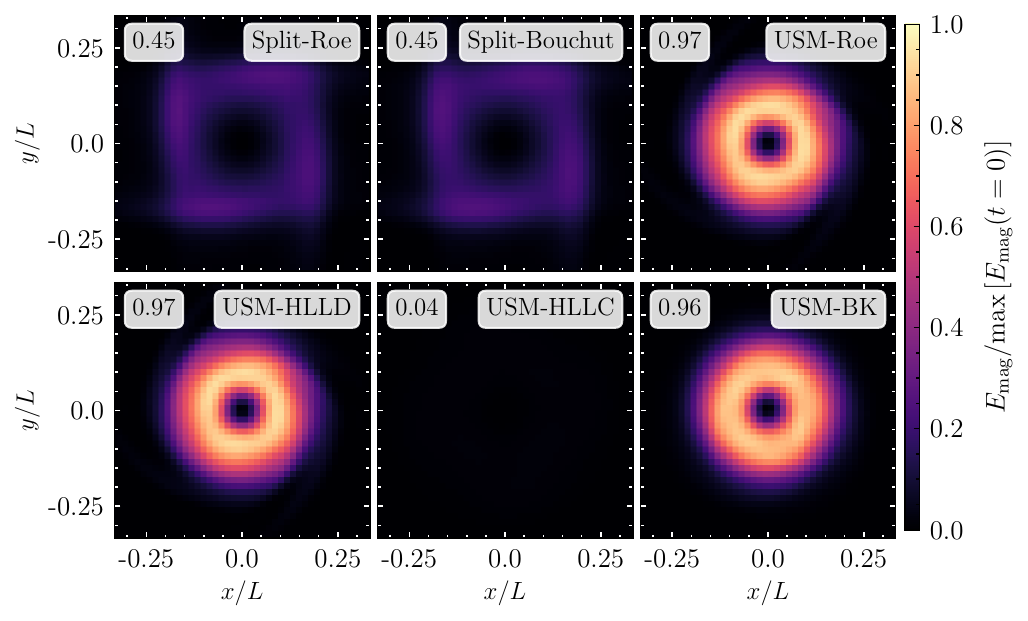}
    \centering
    \caption{Same as Fig.~\ref{fig:LowMachERot}, but for the magnetic energy. The value in the top left corner of each panel shows the fraction of the total magnetic energy left in the system after one complete box advection compared to $t=0$. We find that the USM-BK scheme is also the best-performing scheme with respect to the magnetic energy, with only $4\%$ of the initial energy dissipated. The Split schemes dissipate magnetic energy while damping the magnetic monopoles, while the 3-wave USM-HLLC scheme has dissipated almost all of the magnetic energy in the system.}
    \label{fig:LowMachEMag}
\end{figure}

\begin{figure}
    \centering
    \includegraphics[width=1\linewidth]{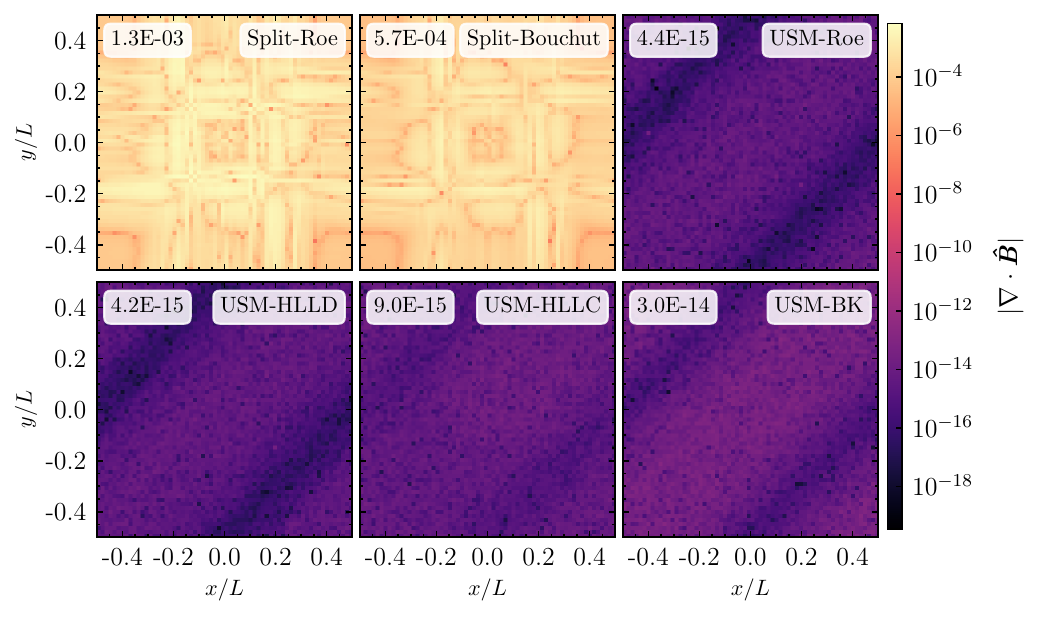}
    \caption{Same as Fig.~\ref{fig:LowMachERot}, but for the divergence of the magnetic field, defined in a normalised fashion via Eq.~(\ref{eq:divb}), such that its magnitude can be compared to order unity. The inset on the top left in each panel shows the root-mean-squared value of $\nabla\cdot\vect{\hat{B}}$. The split schemes keep the value of $\nabla\cdot\vect{B}$ at reasonably low levels while the USM schemes maintain $\nabla\cdot\vect{B}=0$ close to machine precision.}
    \label{fig:DivB}
\end{figure}

In order to quantify the amount of kinetic and magnetic energy dissipation, we calculate and compare the fraction of rotational and magnetic energy that the vortex has retained compared to their respective initial values. The rotational kinetic energy ($E_\mathrm{rot}$) is calculated as
\begin{equation}
    E_\mathrm{rot}=\frac{1}{2}\rho\left[\left(v_\mathrm{x}-\frac{\tilde{v}}{\sqrt{2}}\right)^2 + \left(v_\mathrm{y}-\frac{\tilde{v}}{\sqrt{2}}\right)^2\right],
\end{equation}
while the magnetic energy ($E_\mathrm{mag}$) is calculated as
\begin{equation}
    E_\mathrm{mag}=\frac{1}{2}|\vect{B}|^2.
\end{equation}
Fig.~\ref{fig:LowMachERot} shows the fraction of the rotational energy retained in the system at the end of one complete advection of the vortex. The energy has been normalised by the maximum local rotational energy (at $r=1$) present in the system at the beginning of the evolution. We find that the split schemes (Split-Roe and Split-Bouchut) retain around $65\%$ of the kinetic energy, while the unsplit schemes (USM), except for the USM-BK, retain only $39\%$ of the initial rotational kinetic energy. The new scheme (USM-BK; bottom right panel) performs the best, conserving $88\%$ of the rotational energy.

Fig.~\ref{fig:LowMachEMag} shows the same as Fig.~\ref{fig:LowMachERot}, but for the magnetic energy. Split-Roe and Split-Bouchut lose more than half of their initial magnetic energy and significantly distort the vortex into a nearly square-shaped form. The increased dissipation is a consequence of the divergence-cleaning method, which also dissipates magnetic energy while damping the magnetic monopoles. USM-Roe, USM-HLLD, and USM-BK perform similarly well in conserving the magnetic energy, with USM-BK retaining 96\% of the initial magnetic energy. At the same time, USM-HLLC, which considers only 3 waves in the Riemann solution, dissipates almost all the magnetic energy in the system. All schemes introduce minor distortions in the shape of the vortex. These are much more visible in the split schemes and USM-HLLC, but are minor in USM-Roe, USM-HLLD and USM-BK.

Finally, we look at the divergence of the magnetic field. We define a normalised version of $\nabla\cdot\vect{B}$, as
\begin{equation} \label{eq:divb}
\nabla\cdot\vect{\hat{B}}=\nabla\cdot\frac{\vect{B}\Delta x}{B_{\mathrm{rms}}},
\end{equation}
where $B_\mathrm{rms}$ is the root-mean-squared magnetic field integrated over the entire volume, and $\Delta x$ is the side length of each grid cell.
Fig.~\ref{fig:DivB} shows $|\nabla\cdot\vect{\hat{B}}|$. The choice of Riemann solver does not play any significant role in constraining the divergence of the magnetic field to zero, however, all simulations using constrained transport (USM) perform much better compared to the divergence cleaning used in the split schemes. This is expected since divergence cleaning schemes do not enforce any particular discretisation of $\nabla\cdot\vect{B}$ to zero. They instead rely on diffusing and damping numerical magnetic mono-poles. On the other hand, constrained transport is designed such that $\nabla\cdot\vect{B}=0$ to machine precision by the construction of a particular stencil chosen to construct $\vect{B}$ from the electric field and to calculate $\nabla\cdot\vect{B}$.

\section{Application to magnetic field amplification in low-Mach turbulence}
\label{sec:Turbulent dynamo}

\subsection{Introduction to the turbulent dynamo}
Magnetic fields play an important role in a wide variety of astrophysical systems, including accretion disks \citep{PennaEtAl2010, Boneva2021}, star formation \citep{Choudhari2015, Federrath2015, ShardaEtAl2021}, galaxies \citep{RuzmaikinEtAl1988, BeckAndWielebinski2013}, and the interstellar medium \citep{Fletcher2011, SetaAndFederrath2022}. The presence of strong magnetic fields is attributed to the amplification of seed fields by \textit{turbulent dynamos}. Turbulent dynamos amplify magnetic fields exponentially over short timescales. This amplification is caused by a sequence of "stretching, twisting, folding, and merging" \citep{SchekochihinEtAl2004, BrandeburgAndSubramanian2005, Federrath2016} of magnetic field lines induced by turbulent motions in the plasma, leading to an increase in the density of magnetic field lines in a fluid packet.

\subsection{Numerical method and setup} \label{sec:turb_method}

We solve Eqs.~(\ref{eq:Compressibility})--(\ref{eq:DivB}) in a periodic 3D box of length $L$, uniformly discretised with a grid of $256^3$~cells. Turbulence is driven stochastically by the Ornstein-Uhlenbeck process \citep{EswaranAndPope1988, FederrathEtAl2010} implemented in the publicly available code \texttt{TurbGen} \citep{FederrathEtAl2022}. The turbulence driving field is constructed here to be purely solenoidal (divergence free), using a Helmholtz decomposition in Fourier space, where we measure wave numbers ($k$) in units of $2\pi/L$ . The driving is constrained to large scales, i.e., $k=[1,3]$, following a parabolic Fourier spectrum, where the peak injection is at $k_\mathrm{turb}=2$ and the driving amplitude falls off smoothly to zero at $k=1$ and $k=3$, respectively, as in previous works \citep[e.g.,][]{FederrathEtAl2021}. Using this turbulence driving method, we adjust the overall amplitude of the driver such that the turbulence reaches a target velocity dispersion $\sigma_v = \mach\cs$ on scale $\ell_\mathrm{turb}=L/k_\mathrm{turb}=L/2$, where $\cs$ is the sound speed and $\mach$ is the target turbulence Mach number. This defines the turbulence turnover timescale as $t_\mathrm{turb} = \ell_\mathrm{turb}/\sigma_v=L/(2\mach\cs)$. Here we study sonic Mach numbers of $\mach=0.1$ and $0.01$.

The box is initialised with a uniform density of fluid at rest and the sound speed is set to $\cs=1$, i.e., all speeds are measured relative to the sound speed. The strength and statistical properties of the turbulent dynamo are independent of the structure of the initial magnetic field \citep{SetaAndFederrath2020}, so we initialise a uniform magnetic field in the $z$-direction of the computational domain to obtain a reference Alfv\'en Mach number of $\mathcal{M}_\mathrm{A}=\sigma_v/c_\mathrm{A}=10^{9}$ when the turbulence is fully developed. This corresponds to a very weak initial seed field that is subsequently amplified by the turbulent dynamo.

Finally, for the runs with the USM-BK scheme, we set the cut-off Mach number (see Eq.~\ref{eq:Mcut}) to $M_\mathrm{cut}=\mathcal{M}$. Using $M\mathrm{cut}/\mathcal{M}=1$ is acceptable for chaotic problems like turbulence, where, unlike the Balsara vortex, there is no strict structural symmetry to be preserved.

\subsection{Results for Mach 0.1}
\label{sec:Turb dyn with num dissip}

In ideal-MHD, we set $\nu$ and $\eta$ in Eqs.~(\ref{eq:Compressibility})--(\ref{eq:DivB}) to 0. However, as shown earlier, numerical dissipation is always present owing to finite cell discretisation \citep{ShivakumarAndFederrath2023} and due to the numerical scheme. Consequently, for excessively dissipative solvers, the results from numerical simulations can deviate significantly from the physical setting. In the following sections, we compare the effect of MHD solvers on the time evolution and morphology of the system and calculate the characteristic wave-numbers associated with numerical viscosity and resistivity at Mach~0.1. We run our simulations for a period of $100\,\tturb$ to allow the magnetic field to saturate. However, we focus most of our analyses on the so-called `kinematic phase', where the field does not have a strong back-reaction on the flow yet, and the field grows exponentially fast.

\subsubsection{Time evolution}
\label{sec:Time evolution}

\begin{table}
    \centering
    \captionsetup{type=figure}
    \renewcommand{\arraystretch}{0.01}
    \begin{tabular}{r}
        \includegraphics[width=0.937\linewidth]{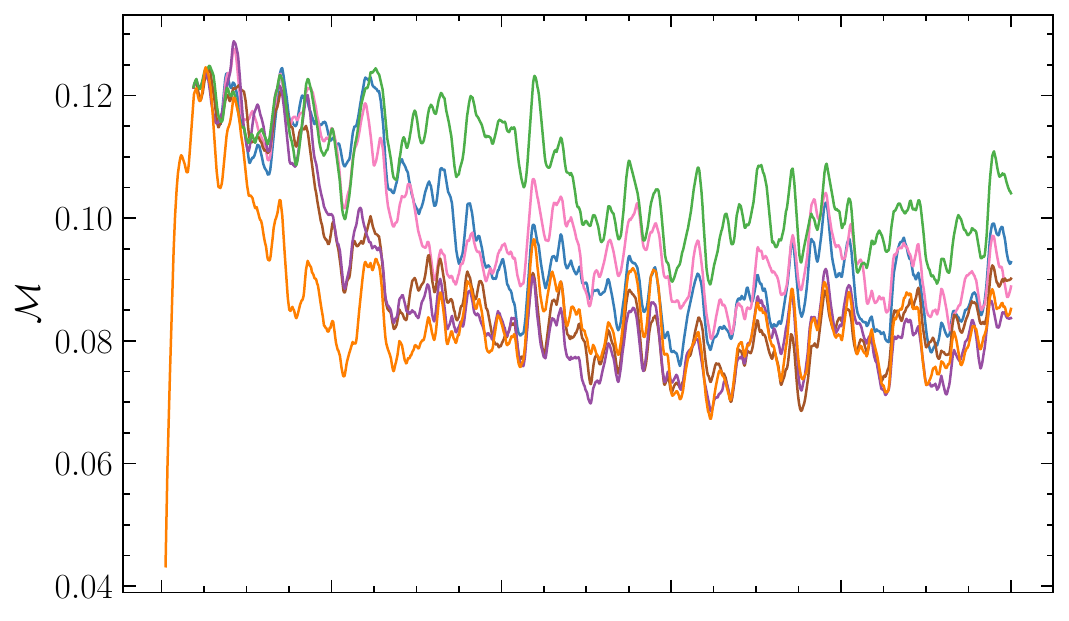} \\[-2.7pt]
        \includegraphics[width=0.946\linewidth]{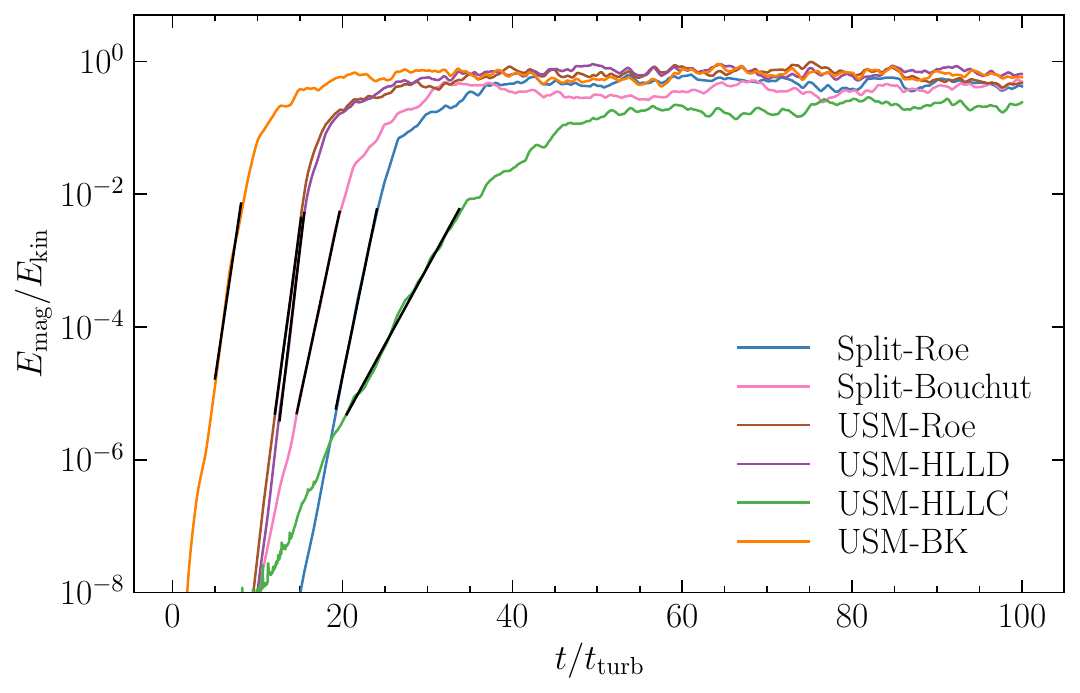}
    \end{tabular}
    \caption{Time evolution of the sonic Mach number (top panel) and the ratio of magnetic to kinetic (turbulent) energy (bottom panel). The Mach number reaches the target value of $\approx0.1$ within $2\tturb$, and during $3$ and $10-30 \tturb$ (depending on the solver; labeled in the legend), the magnetic energy grows exponentially (see fitted solid lines in the bottom panel). Finally, once $\eratio\gtrapprox0.5$, the field saturates and the growth stops, coinciding with a $\approx20\%$ drop in the Mach number (see top panel), due to the enhanced back-reaction of the field onto the flow. USM-HLLD, USM-Roe and USM-BK have the highest growth rate (see Table~\ref{tab:Turbulent dynamo fit parameters}) since they are less dissipative. On the other hand, the 3-wave USM-HLLC has the smallest growth rate, and it shows the weakest dip in Mach number owing to its excessive dissipation of the magnetic energy.}
    \label{fig:Dynamo Growth}
\end{table}

Fig.~\ref{fig:Dynamo Growth} shows the growth of the sonic Mach number ($\mathcal{M}$) in the top panel and the ratio of the magnetic energy to the kinetic energy ($\eratio$) in the bottom panel. We see that the Mach number reaches the target value of $0.1$ within $2\,\tturb$. It is followed by the kinematic phase, where the magnetic energy increases exponentially (up to $10-30\,\tturb$, depending on the numerical scheme used). This is attributed to the turbulent motions of the fluid, which stretch, twist, fold, and merge the magnetic field lines, leading to an increase in their concentration. Finally, as the magnetic field strength increases, the Lorentz force back-reacts on the turbulent motion, suppressing further amplification and saturating the magnetic field. This back-reaction also lowers the Mach number by about 20\%.

In order to measure the magnetic field dynamo growth rate, we fit the exponential model
\begin{equation}
    \frac{E_\mathrm{mag}}{E_\mathrm{kin}}=Ae^{\Gamma t},
    \label{eq:exponential model}
\end{equation}
in the kinematic phase, which we define as $5\times10^{-6}\le\eratio\le5\times10^{-3}$ (i.e., $E_\mathrm{mag}\ll E_\mathrm{kin}$), and $\Gamma$ is the growth rate measured in units of $\tturb^{-1}$. The growth rates measured from the fits are listed in Table~\ref{tab:Turbulent dynamo fit parameters}.

The USM-HLLC scheme exhibits an abnormally low growth rate. Consequently, the magnetic field and the Lorentz force is weaker compared to the other solvers and the sonic Mach number is higher. It also has a lower saturation level (see column~3 in Table~\ref{tab:Turbulent dynamo fit parameters}). This behaviour is attributed to the excessive dissipation of magnetic energy by the HLLC solver (see bottom-middle panel in Fig.~\ref{fig:LowMachEMag}). In contrast, USM-HLLD, USM-Roe and USM-BK achieve the highest growth rates due to their reduced numerical dissipation. It is important to note that the growth rate depends on the magnetic Prandtl number \citep[see][]{FederrathEtAl2014}, which, in turn, is determined by the ratio of the resistive to viscous dissipation wave-numbers (see Appendix~\ref{appendix:Pm k_eta rel}). Therefore, the growth rate is not a universal indicator of solver performance. For instance, USM-HLLD and UMS-Roe simulations exhibit higher effective Prandtl numbers (see Table~\ref{tab:Turbulent dynamo M0.01 effective flow numbers}), resulting in a slightly higher growth rate than that of USM-BK.

\subsubsection{Magnetic field structure}
\label{sec:Magnetic energy morphology}

\begin{figure*}
    \centering
    \includegraphics[width=\linewidth]{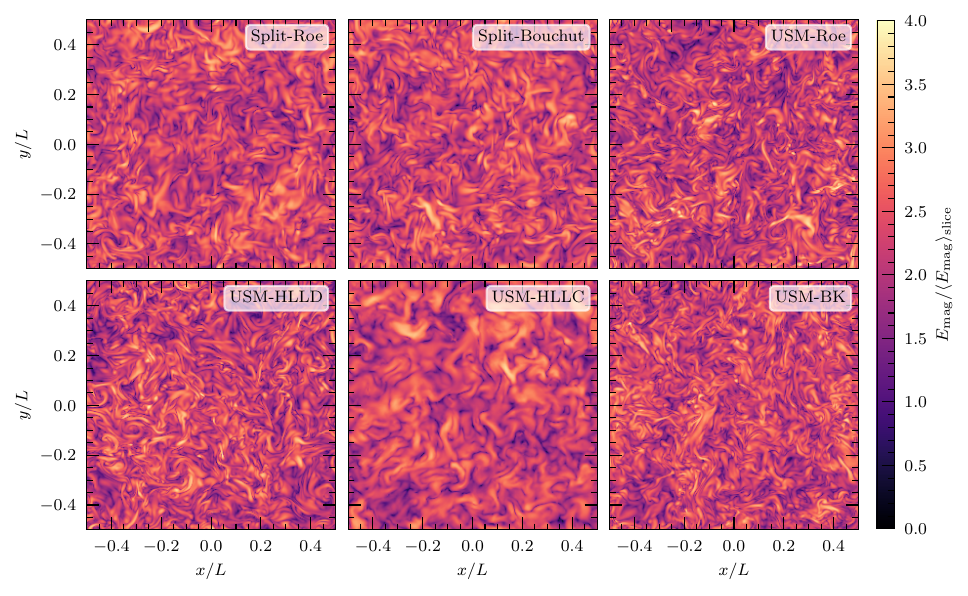}
    \caption{A slice of the magnetic energy normalised by the mean magnetic energy during the kinematic phase of the dynamo, when $\eratio=10^{-4}$, emphasising its spatial distribution. The more dissipative Split schemes and USM-HLLC smear features over large-length scales while USM-Roe, USM-HLLD and USM-BK capture finer structures.}
    \label{fig:Magnetic energy morphology}
\end{figure*}

Numerical dissipation also affects the morphology of the system. To get a qualitative idea of this in turbulent flows, we investigate the spatial distribution of the magnetic energy. Fig.~\ref{fig:Magnetic energy morphology} shows a slice of the magnetic energy normalised by the mean magnetic energy during the kinematic phase of the dynamo, when $\eratio=10^{-4}$. We see random fluctuations in the magnetic energy field with all the solvers, however, the morphology is markedly different in USM-HLLC (bottom-middle panel) and slightly different for Split-Roe and Split-Bouchut (first two panels). They smear the over-densities and the under-densities in the field over larger regions. As mentioned earlier, the dissipation in the Split schemes is attributed to divergence-cleaning, while that in USM-HLLC is a result of its consideration of fewer waves in the Riemann solution. In contrast, USM-HLLD and USM-BK display fine, small-scale structures. We quantitatively analyse the differences between the various schemes in the next section.

\subsubsection{Spectral analysis}
\label{sec:spectral analysis mach 0.1}

\begin{table}
    \centering
    \captionsetup{type=figure}
    \renewcommand{\arraystretch}{0.01}
    \begin{tabular}{r}
        \includegraphics[width=0.95\linewidth]{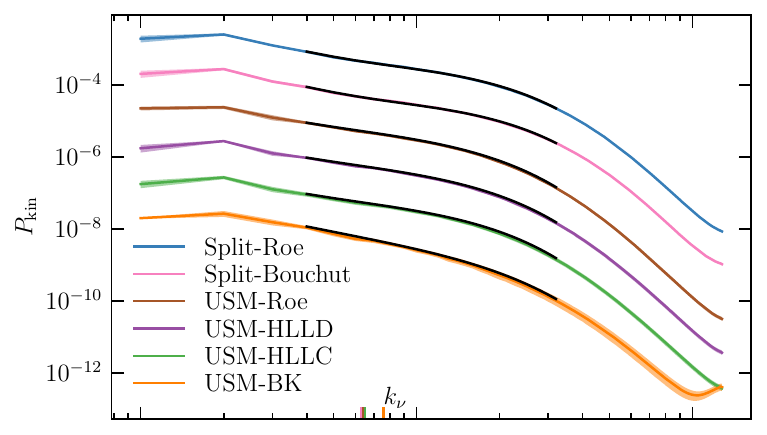}   \\
        \includegraphics[width=0.942\linewidth]{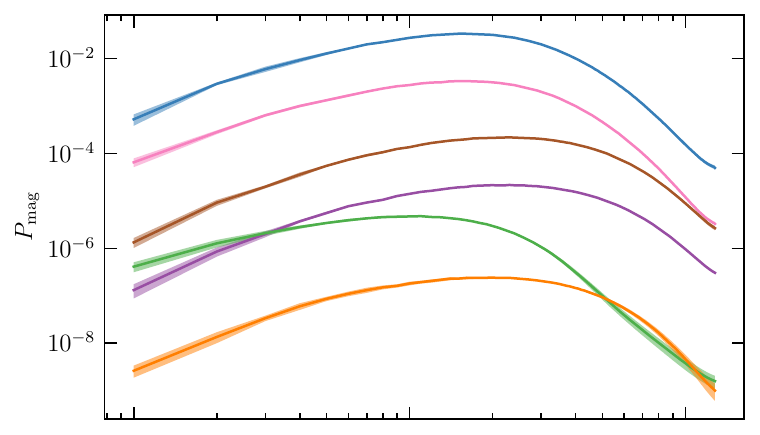} \\
        \includegraphics[width=0.95\linewidth]{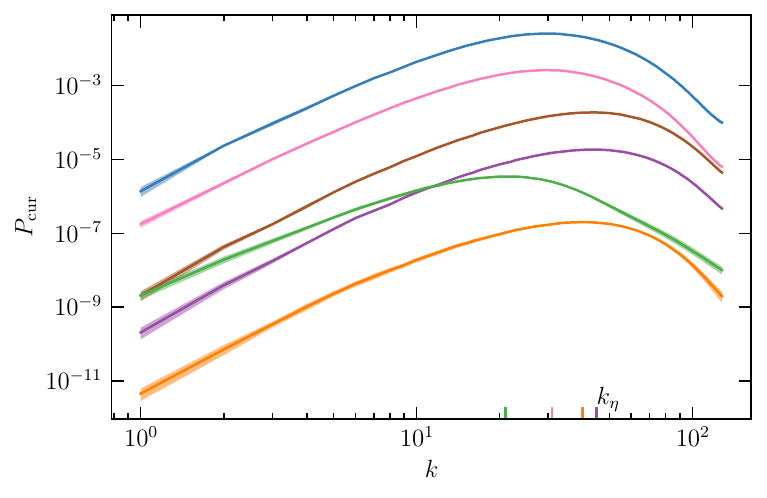}
    \end{tabular}
    \caption{Time-averaged kinetic power spectra (top panel), magnetic power spectra (middle panel), and current power spectra (bottom panel) for various solvers. The spectra are multiplied for every solver except Split-Roe by a factor of 0.1 relative to the next solver in the legends. The black lines in the kinetic power spectra are the fits to the model given in Eq.(~\ref{eq:Pkin}). The viscous dissipation scale and resistive dissipation scale are marked on the x-axis. The viscous dissipation scales are clustered around each other for all solvers, however, the resistive dissipation scale measurements clearly show that HLLC is not suitable for MHD simulations since it is dissipative at very large length scales.}
    \label{fig:Spectra}
\end{table}

\begin{table*}
    \centering
    \caption{Mach~0.1 turbulent dynamo measurements.}
    \setlength{\tabcolsep}{1.8pt}
    \renewcommand{\arraystretch}{1.5}
    \begin{tabular}{lcccccccccc}
    \hline
    Name & $\Gamma(\tturb^{-1})$ & $(E_\mathrm{mag}/E_\mathrm{kin})_\mathrm{sat}$ & $p_\mathrm{bn}$ & $k_\mathrm{bn}$ & $\tilde{k}_\mathrm{\nu}$ & $p_\mathrm{\nu}$ & ${k}_\mathrm{\nu}$ & $k_\mathrm{\eta}$\\
    (1) & (2) & (3) & (4) & (5) & (6) & (7) & (8) & (9)\\
    \hline
    
    Split-Roe & $1.43^{+0.01}_{-0.01}$ & $0.48^{+0.07}_{-0.07}$ & $0.92^{+0.09}_{-0.07}$ & $5.5^{+0.3}_{-0.3}$ & $5.9^{+0.1}_{-0.2}$ & $1.0^{+0.1}_{-0.1}$ & $6.3^{+0.1}_{-0.2}$ & $31^{+1}_{-1}$ \\ 
    Split-Bouchut & $1.39^{+0.01}_{-0.01}$ & $0.40^{+0.06}_{-0.06}$ & $1.01^{+0.10}_{-0.08}$ & $5.8^{+0.3}_{-0.3}$ & $5.9^{+0.1}_{-0.2}$ & $1.0^{+0.1}_{-0.1}$ & $6.3^{+0.1}_{-0.2}$ & $31^{+1}_{-2}$ \\ 
    USM-Roe & $2.20^{+0.01}_{-0.01}$ & $0.66^{+0.13}_{-0.13}$ & $0.42^{+0.28}_{-0.24}$ & $4.0^{+0.8}_{-0.7}$ & $6.8^{+1.7}_{-1.9}$ & $1.0^{+0.1}_{-0.1}$ & $6.4^{+0.6}_{-0.9}$ & $44^{+2}_{-1}$ \\ 
    USM-HLLD & $2.48^{+0.01}_{-0.01}$ & $0.69^{+0.09}_{-0.09}$ & $0.35^{+0.35}_{-0.29}$ & $3.9^{+0.7}_{-0.6}$ & $7.1^{+2.2}_{-2.7}$ & $1.0^{+0.1}_{-0.1}$ & $6.5^{+0.6}_{-1.2}$ & $45^{+1}_{-2}$ \\ 
    USM-HLLC & $0.54^{+0.01}_{-0.01}$ & $0.21^{+0.04}_{-0.04}$ & $0.36^{+0.38}_{-0.29}$ & $4.1^{+1.0}_{-0.7}$ & $7.4^{+2.1}_{-2.5}$ & $1.1^{+0.1}_{-0.1}$ & $6.5^{+0.7}_{-1.0}$ & $21^{+1}_{-2}$ \\ 
    USM-BK & $2.00^{+0.01}_{-0.01}$ & $0.58^{+0.10}_{-0.10}$ & $-0.13^{+0.37}_{-0.30}$ & $4.5^{+1.5}_{-1.0}$ & $8.8^{+2.4}_{-2.8}$ & $1.1^{+0.1}_{-0.1}$ & $7.6^{+1.0}_{-1.1}$ & $40^{+1}_{-0}$ \\     \hline
    \end{tabular}
    \begin{flushleft}
    \textbf{Notes.} All parameters except the saturation value of the ratio of the magnetic energy to the kinetic energy (column 3) were measured/derived by averaging over the kinematic phase of the dynamo when $5\times10^{-6}\le E_\mathrm{mag}/E_\mathrm{kin} \le 5\times10^{-3}$. Columns: (1) Name of the numerical scheme as described in Table~\ref{tab:solvers}. (2) Growth rate in units of $t_\mathrm{turb}^{-1}$. (3) Average value of the ratio of the magnetic energy to the kinetic energy in the saturation phase of the dynamo $(t>60t_\mathrm{turb})$. (4) Exponent of the bottleneck effect in the kinetic spectrum. (5) Scaling wave-number of the bottleneck effect. (6) Viscous dissipation wave-number if $p_\mathrm{\nu}=1$. (7) Exponent of the dissipation term of $P_\mathrm{kin}$. (8) Viscous dissipation wave-number. (9) Resistive dissipation wave-number.        
    \end{flushleft}
    \label{tab:Turbulent dynamo fit parameters}
\end{table*}

In subsonic turbulence, energy cascades from larger scales to smaller scales until it reaches a scale where it is dissipated due to the effects of viscosity and resistivity \citep[e.g.,][]{Frisch1995}. This takes place through the breaking-up of large eddies into smaller eddies. The wave-numbers where viscosity and resistivity act are called viscous dissipation wave-number $k_\nu$, and resistive dissipation wave-number $k_\eta$, respectively.

We calculate the power spectrum of the kinetic energy averaged over the kinematic phase (as defined in Section~\ref{sec:Time evolution}) to measure the viscous dissipation wave-number. We follow the power spectrum model used in \citet{ShivakumarAndFederrath2023} and fit the kinetic spectrum from $k\ge3$ to exclude the turbulence driving scales. The upper limit of the fit is set to $k_\mathrm{max}=N/8=32$, where $N$ is the number of grid cells, to exclude spurious effects that arise on scales smaller than a few grid cells.

The kinetic energy power spectrum ($P_\mathrm{kin}$) in the subsonic regime is modelled as
\begin{equation}
    P_\mathrm{kin}(k)=A_\mathrm{kin}\left[\left(\frac{k}{k_{\mathrm{bn}}}\right)^{-1.7}+\left(\frac{k}{k_{\mathrm{bn}}}\right)^{p_{\mathrm{bn}}}\right]\exp\left[-\left(\frac{k}{\tilde{k}_\mathrm{\nu}}\right)^{p_\mathrm{\nu}}\right],
    \label{eq:Pkin}
\end{equation}
where $A_\mathrm{kin}$ is the amplitude, $k_\mathrm{bn}$ is the scale of energy accumulation due to the bottleneck effect \citep{Falkovich1994, Frisch1995, SchmidtAndHillebrandt2004, VermaAndDiego2007}, $p_\mathrm{bn}$ characterises the strength of the bottleneck effect, and $p_\nu$ characterises the sharpness of the transition into dissipation. The viscous dissipation wave-number as defined in \citet{KrielEtAl2022} is related to $\tilde{k}_\mathrm{\nu}$ and $p_\mathrm{bn}$ by
\begin{equation}
    k_\mathrm{\nu}=\tilde{k}_\mathrm{\nu}^{1/p_\mathrm{\nu}}.
\end{equation}
We point the reader to \citet{ShivakumarAndFederrath2023} and references therein for the motivation behind this model.

To find the characteristic resistive dissipation wave-number ($k_\eta)$, we follow the definition in \citet{KrielEtAl2023}, using the electric current ($\sim\nabla\times\vect{B}$) power spectrum. Since Ohmic dissipation is proportional to current, $k_\eta$ is defined as the wave-number where the current attains a maximum.

The power spectra of kinetic energy, magnetic energy and current are shown in Fig.~\ref{fig:Spectra}, and the fit parameters and the measured characteristic dissipation wave-numbers ($k_\nu$ and $k_\eta$) are given in Table~\ref{tab:Turbulent dynamo fit parameters}. For the kinetic spectra, the dissipation scales are similar for all solvers except USM-BK, which shows dissipation at larger wave-numbers (an $\approx17\%$ difference compared to USM-HLLD). For the current spectra, we see that USM-HLLD marginally outperforms USM-BK (an $\approx12\%$ difference). We also see that the dissipation scale for HLLC lies at very small wave-numbers, i.e., it induces numerical dissipation effects at much larger lengths scales, smearing out small-scale features. Thus, it is particularly unsuitable for modelling MHD flows.

\subsubsection{Numerical Reynolds numbers}

\begin{table}
    \centering
    \caption{Mach~0.1 turbulent dynamo effective Reynolds numbers.}
    \setlength{\tabcolsep}{1.8pt}
    \renewcommand{\arraystretch}{1.5}
    \begin{tabular}{lcccc}
    \hline
    Name          & $\mathrm{Re}$     & $\mathrm{Rm}$     & $\mathrm{Pm}$ \\
    (1)           & (2)               & (3)               & (4)           \\
    \hline
    Split-Roe & $6.3^{+2.9}_{-1.5}\times10^2$ & $3.0^{+2.5}_{-1.5}\times10^3$ & $4.7^{+2.9}_{-2.2}$\\ 
    Split-Bouchut & $6.4^{+2.7}_{-1.5}\times10^2$ & $3.0^{+2.5}_{-1.4}\times10^3$ & $4.6^{+2.9}_{-2.0}$\\ 
    USM-Roe & $6.4^{+2.9}_{-1.8}\times10^2$ & $6.3^{+5.4}_{-3.0}\times10^3$ & $9.9^{+7.0}_{-4.6}$\\ 
    USM-HLLD & $6.4^{+3.3}_{-2.0}\times10^2$ & $6.1^{+5.1}_{-3.0}\times10^3$ & $9.4^{+7.3}_{-4.4}$\\ 
    USM-HLLC & $6.5^{+3.2}_{-1.9}\times10^2$ & $1.4^{+1.2}_{-0.7}\times10^3$ & $2.1^{+1.6}_{-1.0}$\\ 
    USM-BK & $8.1^{+4.0}_{-2.4}\times10^2$ & $4.6^{+3.7}_{-2.2}\times10^3$ & $5.6^{+4.2}_{-2.6}$\\ 
    \hline
    \end{tabular}
    \label{tab:Turbulent dynamo effective flow numbers M0.1}
\end{table}

Since numerical dissipation is always present in MHD simulations, the simulated flows have a finite numerical hydrodynamic Reynolds number (Re) and magnetic Reynolds number (Rm), in contrast to a perfectly ideal setting, where these would be infinite, because $\nu=\eta=0$ in the MHD equations. Appendix~\ref{appendix:Pm k_eta rel} describes the procedure for obtaining Re and Rm from the characteristic dissipation scales, using the key relations,
\begin{equation}
    \mathrm{Re}=\left(\frac{k_\nu}{c_\mathrm{Re} k_\mathrm{driving}}\right)^{4/3},
\end{equation}
\begin{equation}
    \mathrm{Pm}=\left(\frac{k_\mathrm{\eta}}{c_\mathrm{Pm}k_\mathrm{\nu}}\right)^2,\;\mathrm{and}
\end{equation}
\begin{equation}
    \mathrm{Rm}=\mathrm{Re}\times\mathrm{Pm},
\end{equation}
where $c_\mathrm{Re}=0.025^{+0.005}_{-0.006}$ and $c_\mathrm{Pm}=2.3^{+0.8}_{-0.5}$.

Table~\ref{tab:Turbulent dynamo effective flow numbers M0.1} lists the values of the effective Re, Rm, and Pm for the Mach~0.1 simulations, for each numerical scheme. The measured values of Re range between $630^{+290}_{-150}$ and $810^{+400}_{-240}$ (an $\approx30\%$ variation), while Rm values vary between $1400^{+1200}_{-700}$ and $6300^{+5400}_{-3000}$ (a striking $\approx350\%$ variation attributable to the choice of the numerical scheme). The Pm values (which can be calculated from Re and Rm) vary between $2.1^{+1.6}_{-1.0}$ and $9.9^{+7.0}_{-4.6}$ (an $\approx370\%$ variation). We see that the USM-BK scheme shows the highest Re ($810^{+400}_{-240}$ compared to $640^{+330}_{-200}$ for USM-HLLD, the next-highest value). This suggests that USM-BK has the least dissipation of kinetic energy in the low-Mach regime among the solvers/schemes compared; however, the large error bars show that the difference between the dissipation properties of the USM-HLLD and USM-BK is not statistically significant at Mach~0.1. We further find that USM-HLLD and USM-Roe have a higher Pm than USM-BK ($9.4^{+7.3}_{-4.4}$ and $9.9^{+7.0}_{-4.6}$ for USM-HLLD and USM-BK, respectively, compared to $5.6^{+4.2}_{-2.6}$ for USM-BK). As the dynamo growth rate depends on Re as well as Pm \citep[see][]{FederrathEtAl2014}, these measurements of Pm explain why USM-BK has a somewhat smaller growth rate (c.f., Fig.~\ref{fig:Dynamo Growth} and Tab.~\ref{tab:Turbulent dynamo fit parameters}) compared to USM-HLLD, despite being the least dissipative solver.

\subsection{Results for Mach 0.01}

We have already established in the Balsara vortex test (c.f., Sec.~\ref{sec:balsara}) that the schemes using Dedner-Marder cleaning do not perform very well in constraining the divergence of the magnetic field to zero. From our analysis of the electric current power spectra, it is quite clear that HLLC is not suitable for low-Mach simulations due to its large resistive dissipation length scale. Based on our study of energy conservation and the current power spectra, we can conclude that USM-HLLD and USM-BK have been the best-performing solvers so far, showing comparable results. Given that many astrophysical processes, such as stratified stellar flows \citep[see][]{KupkaAndMuthsam2017} and early-Universe turbulent dynamos \citep[see][]{ChirakkaraEtAl2021}, involve highly subsonic flows with Mach numbers below $10^{-2}$, we test these schemes further by running a turbulent dynamo simulation at Mach~0.01. Since our focus is on the kinematic stage of the dynamo (which we use to obtain solver properties), we stop our simulations close to the onset of saturation.

\subsubsection{Time evolution}

\begin{figure}
    \centering
    \includegraphics[width=1\linewidth]{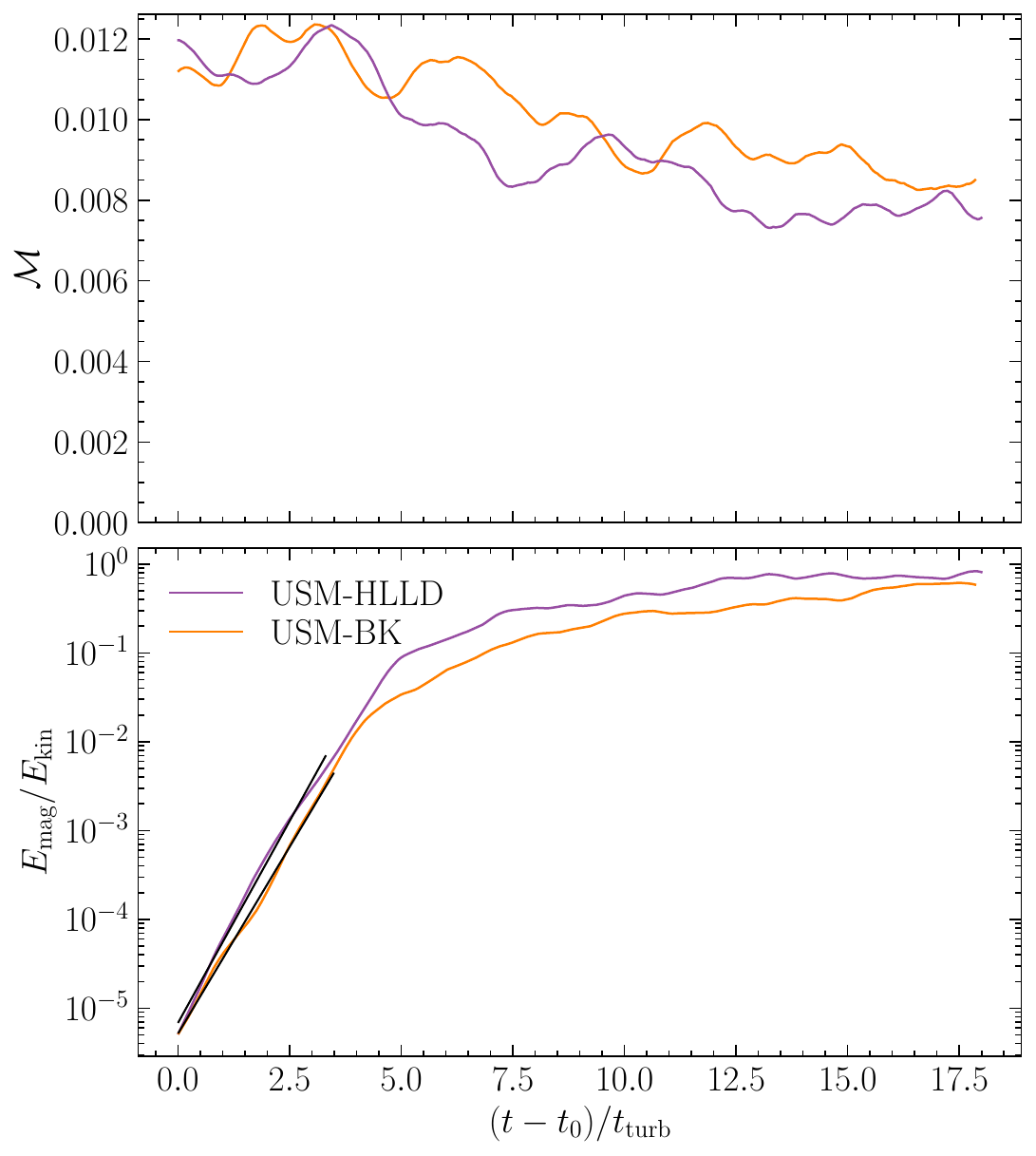}
    \caption{Similar to Fig.~\ref{fig:Dynamo Growth}, but for Mach~0.01, comparing the best-performing solvers from the previous comparison at Mach~0.1, namely USM-BK (orange) and USM-HLLD (purple). Note that the x-axis shows $(t-t_0)/\tturb$, where $t_0$ is chosen such that both runs start at the same $\eratio$ to facilitate the comparison.}
    \label{fig:Dynamo Growth M0.01}
\end{figure}

Fig.~\ref{fig:Dynamo Growth M0.01} shows the evolution of the Mach number and the ratio of magnetic energy to kinetic energy with time. The plots have been shifted so that both simulations have the same starting ratio of the kinetic energy to the magnetic energy, facilitating the comparison, as the initial conditions are not relevant for the turbulent dynamo \citep{SetaAndFederrath2020,BeattieEtAl2023}. The features are similar to what we see at Mach~0.1 (cf., Fig.~\ref{fig:Dynamo Growth}). We define the kinematic phase as in Section~\ref{sec:Time evolution} ($5\times10^{-6}\leq\eratio\leq5\times10^{-3}$), with the measured growth rate listed in Table~\ref{tab:Turbulent dynamo M0.01 fit parameters}. The growth rate is slightly higher for USM-HLLD compared to USM-BK, which is consistent with the higher Prandtl number of USM-HLLD (see Table~\ref{tab:Turbulent dynamo M0.01 effective flow numbers}), similar to what we found for the Mach~0.1 comparison of the two solvers.

\subsubsection{Morphology}

\begin{figure*}
    \centering
    \setlength{\tabcolsep}{-4.6pt}
    \begin{tabular}{ccc}
        \includegraphics[height=0.526\textheight]{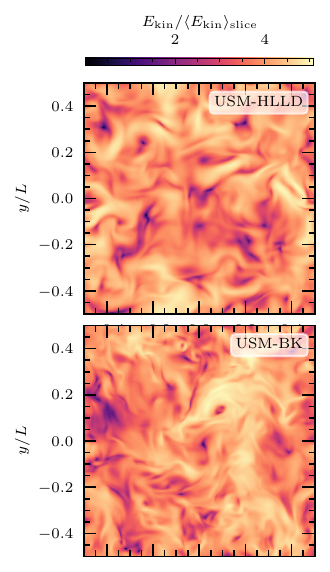} &
        \includegraphics[height=0.526\textheight]{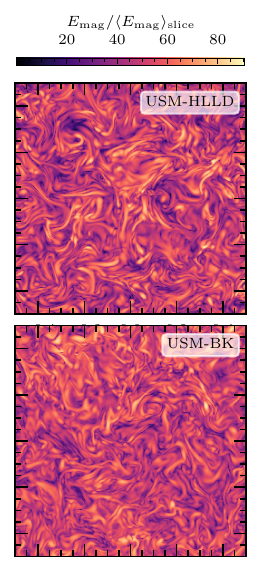} &
        \includegraphics[height=0.526\textheight]{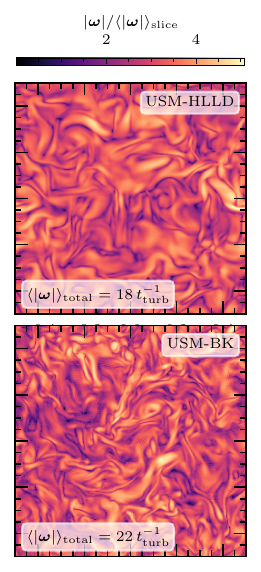}
    \end{tabular}
    \caption{Slices of kinetic energy (left), magnetic energy (middle), and vorticity, $\vect{\omega}=\nabla\times\vect{v}$ (right), through the simulation domain taken during the kinematic phase of the turbulent dynamo at Mach~0.01 when $\eratio=10^{-4}$. The presence of small-scale structure in USM-BK shows that it dissipates kinetic energy at smaller length scales compared to USM-HLLD. Compared to the kinetic energy, the magnetic energy (middle panels) shows somewhat smaller qualitative differences between the two solvers, but it appears that also here the USM-BK captures slightly more small-scale turbulent structure than USM-HLLD; quantified in Section~\ref{sec:M 0.01 spectral analysis}. Finally, the vorticity (right-hand panels) reinforces the finding that USM-BK captures more small-scale structure than USM-HLLD. The inset labels on the vorticity panels show measurements of the mean vorticity in the entire system (not just the slice), demonstrating that USM-BK captures $\sim20\%$ more vorticity than USM-HLLD.}
    \label{fig:Morphology M0.01}
\end{figure*}

Fig.~\ref{fig:Morphology M0.01} shows the kinetic energy, the magnetic energy, and the vorticity, respectively, in a slice during the kinematic phase when $\eratio=10^{-4}$. We see that more small-scale kinetic structure is captured in USM-BK compared to USM-HLLD. This is consistent with the fact that USM-BK dissipates kinetic energy at smaller length scales compared to USM-HLLD. The presence of small-scale structures (left panel) shows that smaller eddies are present in the USM-BK test case, whereas USM-HLLD dissipates energy into heat before forming eddies of comparable sizes. A similar pattern is hinted by the magnetic energy (middle panel), where USM-BK captures somewhat more small-scale structure than USM-HLLD. This difference is demonstrated quantitatively in the next section. Our findings are further corroborated by the vorticity modulus (right panel), where USM-BK captures $\sim20\%$ more vorticity (see inset label) than USM-HLLD.

\subsubsection{Spectral analysis}
\label{sec:M 0.01 spectral analysis}

\begin{figure}
    \centering
    \includegraphics[width=0.98\linewidth]{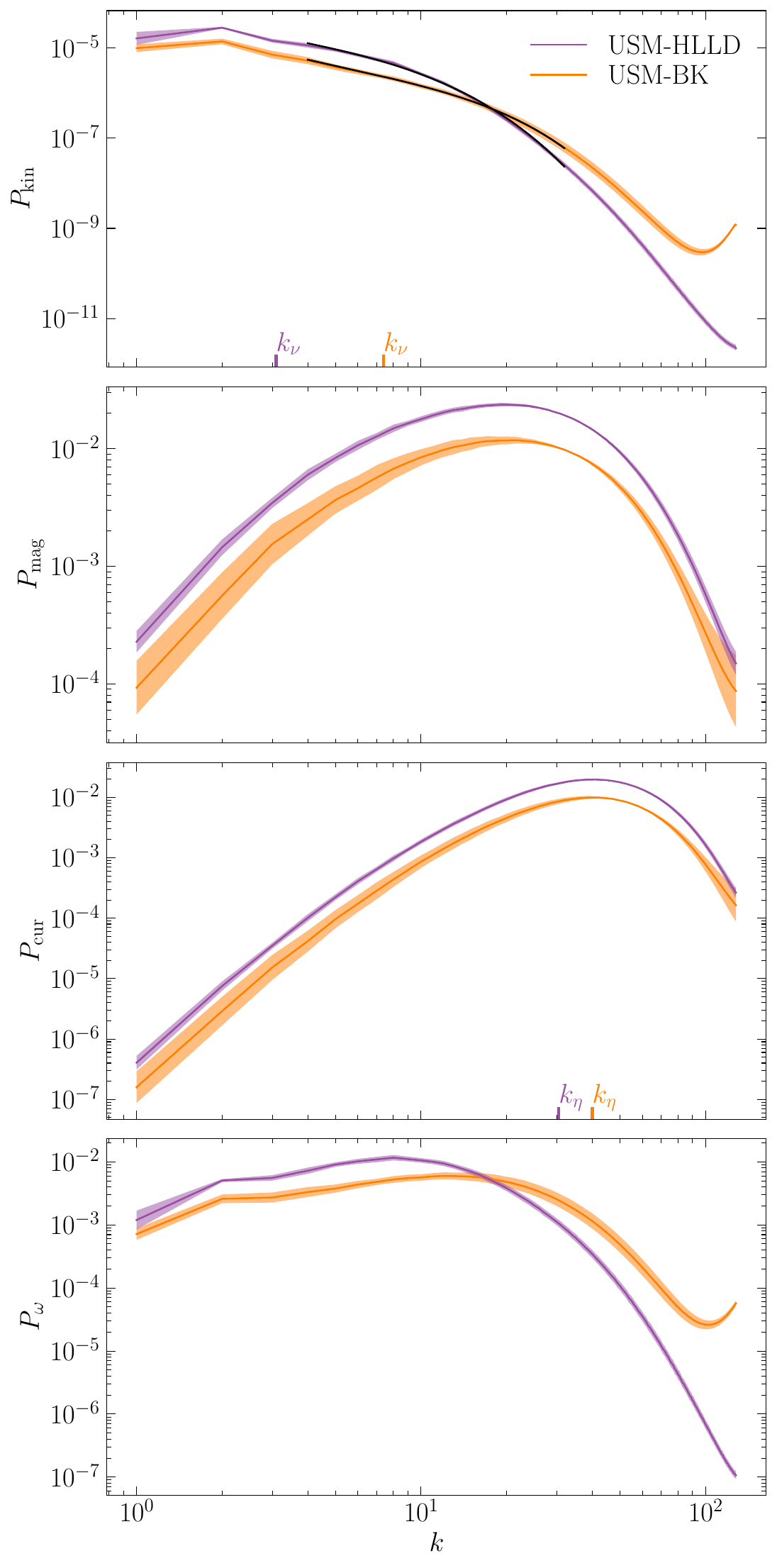}
    \caption{Time-averaged power spectra of the kinetic energy (top panel), magnetic energy (second panel), electric current (third panel) and vorticity (bottom panel), for USM-HLLD and USM-BK. USM-BK has been shifted by a factor of 0.5 along the y-axis for clarity. The black lines are the fits to the model given in Eq.~(\ref{eq:Pkin}). The viscous dissipation scale and resistive dissipation scale are marked on the x-axis. Both the viscous and resistive dissipation scales are significantly separated from each other, showing that USM-BK is less dissipative than USM-HLLD, in both kinetics and magnetics.}
    \label{fig:Spectra M0.01}
\end{figure}

\begin{table*}
    \centering
    \caption{Same as Tab.~\ref{tab:Turbulent dynamo fit parameters}, but for Mach~0.01. \label{tab:Turbulent dynamo M0.01 fit parameters}}
    \setlength{\tabcolsep}{1.8pt}
    \renewcommand{\arraystretch}{1.5}
    \begin{tabular}{lcccccccccc}
    \hline
    Name & $\Gamma(t_\mathrm{turb}^{-1})$ & $p_\mathrm{bn}$ & $k_\mathrm{bn}$ & $\tilde{k}_\mathrm{\nu}$ & $p_\mathrm{\nu}$ & ${k}_\mathrm{\nu}$ & $k_\mathrm{\eta}$\\
    (1) & (2) & (3) & (4) & (5) & (6) & (7) & (8) \\
    \hline
    USM-HLLD & $2.09^{+0.02}_{-0.02}$ & $0.3^{+0.2}_{-0.2}$ & $2.0^{+0.7}_{-0.4}$ & $2.6^{+0.8}_{-0.6}$ & $0.8^{+0.1}_{-0.1}$ & $3.1^{+0.8}_{-0.7}$ &  $31^{+9}_{-6}$ \\ 
    USM-BK & $1.94^{+0.01}_{-0.01}$ & $-0.2^{+0.2}_{-0.2}$ & $4.7^{+0.5}_{-0.5}$ & $10.6^{+1.4}_{-1.4}$ & $1.2^{+0.1}_{-0.1}$ & $7.4^{+1.0}_{-0.6}$ &  $40^{+5}_{-3}$ \\ 
    \hline
    \end{tabular}
\end{table*}

We repeat the analysis in Section~\ref{sec:spectral analysis mach 0.1} for the two Mach~0.01 runs. Fig.~\ref{fig:Spectra M0.01} shows the kinetic energy, magnetic energy, current, and vorticity power spectra. The kinetic spectra reveal that USM-HLLD turns downwards (a sign of the onset of dissipation) on scales larger (wave-numbers smaller) than USM-BK, implying that USM-BK dissipates kinetic energy at smaller length scales compared to USM-HLLD, and is therefore less dissipative. The current power spectrum peaks at a larger wave-number for USM-BK than USM-HLLD, implying that magnetic resistivity starts acting at smaller length scales for USM-HLLD compared to USM-BK, i.e., USM-HLLD is somewhat more resistive than USM-BK. The sharp downward turn of the vorticity power spectrum of USM-HLLD indicates that smaller eddies have been dissipated into heat, a consequence of smaller viscous dissipations wave-number. The fitted dissipation wave-numbers are reported in Table~\ref{tab:Turbulent dynamo M0.01 fit parameters}. We find that USM-BK dissipates at significantly smaller length scales both in terms of kinetics (58\% difference in the wave-numbers) and magnetics (23\% difference in the wave-numbers).

\subsubsection{Numerical Reynolds numbers}

\begin{table}
    \centering
    \caption{Same as Table~\ref{tab:Turbulent dynamo effective flow numbers M0.1}, but for Mach~0.01.}
    \setlength{\tabcolsep}{1.8pt}
    \renewcommand{\arraystretch}{1.5}
    \begin{tabular}{lcccc}
        \hline
        Name     & $\mathrm{Re}$     & $\mathrm{Rm}$     & $\mathrm{Pm}$ \\
        (1)      & (2)               & (3)               & (4)           \\
        \hline
        USM-HLLD & $2.4^{+1.4}_{-0.8}\times10^2$ & $5.5^{+6.8}_{-3.1}\times10^3$ & $21^{+31}_{-12}$\\ 
        USM-BK & $8.0^{+3.8}_{-2.2}\times10^2$ & $5.2^{+5.7}_{-2.7}\times10^3$ & $6.5^{+6.1}_{-3.3}$\\ 
        \hline
    \end{tabular}
    \label{tab:Turbulent dynamo M0.01 effective flow numbers}
\end{table}

Following the relations given in Appendix~\ref{appendix:Pm k_eta rel}, we measure the numerical hydrodynamic and magnetic Reynolds numbers, and the Prandtl number in Table~\ref{tab:Turbulent dynamo M0.01 effective flow numbers}. We find that USM-BK has $\mathrm{Re}=800^{+380}_{-220}$, while USM-HLLD has $\mathrm{Re}=240^{+140}_{-80}$, implying that the former is less dissipative. We note that the Re for USM-HLLD has dropped by a factor of 2.5, compared to the Mach~0.1 run, while USM-BK has roughly the same value, demonstrating that the solver successfully retains high values of Re even at low Mach number. The Pm for USM-BK is lower $\left(\mathrm{Pm}=6.5^{+6.1}_{-3.3}\right)$ compared to USM-HLLD $\left(\mathrm{Pm}=21^{+31}_{-12}\right)$, which explains why USM-BK has a lower growth rate in Table~\ref{tab:Turbulent dynamo M0.01 fit parameters}. Since Pm is so much higher in USM-HLLD, the product of Re and Pm, i.e., Rm, turns out to be similar for both solvers.

\subsection{Computational Cost}

Factoring out the sound speed ($c_\mathrm{s}$) in Eq.~\eqref{eq:lambda_fastest} and approximating the Alfv\'en speed in the x-direction by the total Alfv\'en speed ($c_\mathrm{A;x}\approx c_\mathrm{A}$), we can write the fastest wave-speed as
\begin{equation}
     \lambda_\mathrm{fastest} \approx \cs M +
     \frac{\cs}{2}\sqrt{\left(\frac{1}{M_\mathrm{BK}^2}+\frac{M^2}{M_\mathrm{A}^2}\right) + \sqrt{\left(\frac{1}{M_\mathrm{BK}^2}+\frac{M^2}{M_\mathrm{A}^2}\right)^2-4\frac{M^2}{M_\mathrm{A}}}},
\end{equation}
where $M_\mathrm{BK}=1$ for conventional Riemann solvers, while $M_\mathrm{BK}\approx \mathcal{M}$ for the BK method, for our choice of $M_\mathrm{cut}=\mathcal{M}$.
For our application of turbulent dynamos where $M\approx\mathcal{M}\ll 1\ll \mathcal{M}_\mathrm{A}\approx M_\mathrm{A}$ and $c_\mathrm{s}\approx1$, this can be written as
\begin{equation}
    \lambda_\mathrm{fastest}\sim \mathcal{M}+\frac{1}{M_\mathrm{BK}}\,.
\end{equation}
We can immediately see that for conventional Riemann solvers, the fastest signal speed scales as $(1+\mach)\sim1$ while for the BK method it scales as $\sim1/\mach$. Therefore, the time-step ($\Delta t\propto c_\mathrm{fastest}^{-1}$) is independent of the Mach number for conventional solvers, however, $\Delta t\propto \mach$ for the BK method (see Eq.~\ref{eq:CFL condition}).

\begin{figure}
    \centering
    \includegraphics[width=1\linewidth]{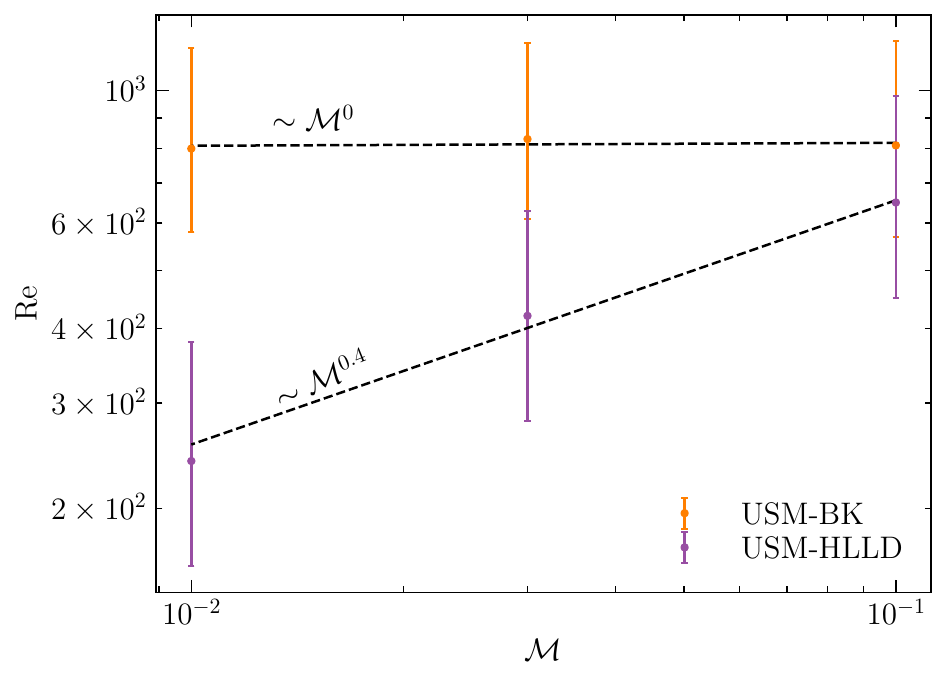}
    \caption{The effective numerical hydrodynamic Reynolds number ($\mathrm{Re}$) as a function of the turbulent Mach number ($\mathcal{M}$). $\mathrm{Re}$ does not change with $\mathcal{M}$ for USM-BK, however, it decreases with decreasing Mach number for USM-HLLD.}
    \label{fig:ReMach}
\end{figure}

Since the turbulent turnover time ($t_\mathrm{turb}$) also scales as $\sim1/\mach$ (c.f., Sec.~\ref{sec:turb_method}), the number of time-steps required to achieve the same amount of time evolution (i.e., the number of eddy-turnover times of evolution) also scales as $\sim1/\mach$. Therefore, the total cost of a simulation with the BK method scales as $1/\mach^2$, while that for conventional Riemann solvers goes as $1/\mach$. If the end goal is to achieve the highest numerical $\mathrm{Re}$ possible, either USM-BK can be used at some resolution or USM-HLLD can be used at a comparatively higher resolution. Figure~\ref{fig:ReMach} shows the variation of numerical $\mathrm{Re}$ with $\mathcal{M}$ at fixed resolution ($N$). For USM-BK, $\mathrm{Re}$ does not change significantly with $\mathcal{M}$, however, it scales as $\mathcal{M}^{0.4}$ for USM-HLLD. \citet{ShivakumarAndFederrath2023} have shown that the effective numerical $\mathrm{Re}$ varies as $N^{4/3}$, where $N$ is the number of grid cells in each direction. 

Therefore, for a given Mach number and resolution, the effective $\mathrm{Re}$ for the two solvers goes as
\begin{eqnarray}
    \mathrm{Re_{BK}}\sim\mathcal{M}^{0}N^{4/3}\,,\nonumber\\
    \mathrm{Re_{HLLD}}\sim\mathcal{M}^{0.4}N^{4/3}\,.
    \label{eq:Re_scaling}
\end{eqnarray}
The ratio of computational cost ($\mathcal{C}$) for the two solvers at a given Mach number and resolution is given by
\begin{equation}
    \frac{\mathcal{C}_\mathrm{BK}}{\mathcal{C}_\mathrm{HLLD}}\sim\frac{\mathcal{M}^{-2}N^4}{\mathcal{M}^{-1}N^4}\,.
    \label{eq:cost_scaling}
\end{equation}
To achieve a target $\mathrm{Re}=\mathrm{Re_{BK}}=\mathrm{Re_{HLLD}}$ at a fixed Mach number by varying the resolution, we can use Eq.~\eqref{eq:Re_scaling} and Eq.~\eqref{eq:cost_scaling}, and write the ratio of the computational cost as
\begin{equation}
    \frac{\mathcal{C}_\mathrm{BK}}{\mathcal{C}_\mathrm{HLLD}}\sim\frac{\mathcal{M}^{-2}\mathrm{Re}^3}{\mathcal{M}^{-1}\mathcal{M}^{-1.2}\mathrm{Re}^3}\sim\mathcal{M}^{0.2}\,.
\end{equation}
Thus, we can say that USM-BK is marginally better than USM-HLLD on the metric of computational cost. An implicit implementation of the BK method, which is less restrictive in the time-step constraint is preferable and we leave it to future works.

% ==== Conclusions ====
\section{Conclusions} \label{sec:Conclusion}

We examined the impact of numerical schemes, particularly the choice of Riemann solver, on numerical dissipation in low-Mach MHD simulations. Using the Balsara vortex test problem, we assessed the suitability of the new Riemann solver (USM-BK) and explored its effectiveness in capturing structures in low-Mach turbulent dynamo simulations. The following are the main conclusions drawn from this work:

\begin{enumerate} 
    \item Conventional Riemann solvers (Roe, HLLC, HLLD, and Bouchut) exhibit excessive dissipation in the low-Mach regime.
    \item The new USM-BK solver demonstrates the least dissipation of kinetic energy in the Balsara vortex tests at Mach~0.01, preserving 84\% of the kinetic energy after one complete advection of the vortex across the computational grid. In constrast, USM-Roe, USM-HLLC, and USM-HLLD show the highest dissipation, preserving only 39\% of the kinetic energy.
    \item The USM-BK solver also exhibits very little dissipation of magnetic energy at Mach~0.01, retaining 96\% of the magnetic energy after one complete advection of the Balsara vortex. It is marginally outperformed by USM-HLLD and USM-Roe, which retain 97\% of the magnetic energy.
    \item An unsplit-staggered mesh (USM) implementation of the BK Riemann solver is preferred over the usage of divergence cleaning since divergence cleaning also diffuses the magnetic energy. Furthermore, constrained transport keeps the divergence of magnetic field close to zero up to machine precision, by construction.
    \item The time-step restriction for stability for USM-BK scales as $\Delta t\sim O(\mathcal{M})$ (c.f., Eq.~\ref{eq:CFL condition}). Therefore, implicit time-steppers are preferable for applications in the low Mach regime.
    \item As discussed in Section~\ref{sec:Turbulent dynamo}, the choice of Riemann solver significantly influences both the growth rate and the saturation level of the dynamo due to variations in the effective Reynolds numbers between solvers. We measured the growth rates and the corresponding effective Reynolds numbers for various solvers at Mach~0.1 and Mach~0.01.
    \item The new solver can resolve smaller length scales compared to the other solvers, which is evident from the kinetic and electric current spectra. While USM-HLLD marginally outperforms USM-BK in resolving magnetic structures at Mach~0.1, USM-BK surpasses USM-HLLD in performance as the Mach number is decreased. This difference in dissipation length scales is also reflected in the morphological features seen in snapshots taken during the kinematic phase.
    \item At a given energy ratio, USM-BK captures more vorticity compared to USM-HLLD. The vorticity power spectra show that USM-BK has more power at smaller wave-numbers than USM-HLLD, indicating that USM-BK captures more small-scale structures since smaller eddies have not been dissipated into heat.
\end{enumerate}

We conclude that the new USM-BK solver is the most suitable for low-Mach MHD simulations, among the schemes compared, as it exhibits the least dissipation of kinetic and magnetic energy. The solver is particularly effective in capturing small-scale structures of the flow, making it a good choice for turbulent dynamo simulations in the low-Mach regime.

\section*{Acknowledgements}
We thank Knut Waagan and Dinshaw Balsara for helpful discussions. C.~F.~acknowledges funding provided by the Australian Research Council (Discovery Project grants~DP230102280 and~DP250101526), and the Australia-Germany Joint Research Cooperation Scheme (UA-DAAD). We further acknowledge high-performance computing resources provided by the Leibniz Rechenzentrum and the Gauss Centre for Supercomputing (grants~pr32lo, pr48pi, pn76ga and GCS Large-scale project~10391), the Australian National Computational Infrastructure (grant~ek9) and the Pawsey Supercomputing Centre (project~pawsey0810) in the framework of the National Computational Merit Allocation Scheme and the ANU Merit Allocation Scheme. The simulation software, \texttt{FLASH}, was in part developed by the Flash Centre for Computational Science at the University of Chicago and the Department of Physics and Astronomy at the University of Rochester.
We also acknowledge the use of OpenAI's ChatGPT for assistance in proofreading and improving the clarity of the manuscript, and GitHub Copilot for assistance in generating code.

%%%%%%%%%%%%%%%%%%%%%%%%%%%%%%%%%%%%%%%%%%%%%%%%%%
\section*{Data Availability}

The data used in this article (approximately 4~TB) is available upon reasonable request to the authors.


\bibliographystyle{mnras}
\bibliography{references}

%%%%%%%%%%%%%%%%% APPENDICES %%%%%%%%%%%%%%%%%%%%%

\appendix

\section{Choice of the cut-off Mach number ($M_\mathrm{cut}$)}
\label{appendix:Mcut}

Fluxes obtained from an approximate Riemann solvers have a dissipation term in the momentum/pressure flux of the form
\begin{equation}
    D \sim \rho \lambda_\mathrm{fastest}(u^\mathrm{L} - u^\mathrm{R}),
\end{equation}
where $\rho$ and $\lambda_\mathrm{fastest}$ are suitable approximations of density and the fastest wave speed at the cell interfaces, and $u^\mathrm{L}$ and $u^\mathrm{R}$ are the x-components (direction along which the MHD equations are one-dimensionalised before solving the Riemann problem) of the fluid speed on the left and right side of the cell interface, respectively. This dissipation term adds artificial viscosity that dampens down any spurious perturbation modes generated in the solution of the discretized MHD equations. However, in the low-Mach regime, such a dissipation term overwhelms the physical flux. This is clear from the ratio of the advective flux to the dissipation term:
\begin{equation}
    \frac{F_\mathrm{advective}}{D} \sim \frac{\rho u^2}{\rho \lambda_\mathrm{{fastest}}\Delta u}\sim\mathcal{M}.
\end{equation}
The advective flux is smaller compared to the dissipation term in subsonic flows. 

This incorrect scaling is fixed by rescaling the dissipation term by a factor $\phi\propto\mathcal{M}$. \citet{BirkeKlingenberg2023} use $\phi=M_\mathrm{BK}$ defined in Eq.~\eqref{eq:Mcut} as
\begin{equation}
    M_\mathrm{BK} = \mathrm{min}\left\{\mathrm{max}\left\{M_\mathrm{cut}, \frac{u}{c_\mathrm{s}}\right\}, 1\right\}.
\end{equation}
$M_\mathrm{cut}$ sets a local cut-off Mach number below which dissipation is no longer reduced. It also controls the fastest wave-speed in the relaxation scheme (see Eq.~\ref{eq:lambda_fastest}). Our relaxation scheme is less diffusive as long as $M_\mathrm{cut} < 1$.

\begin{figure}
    \centering
    \includegraphics[width=1\linewidth]{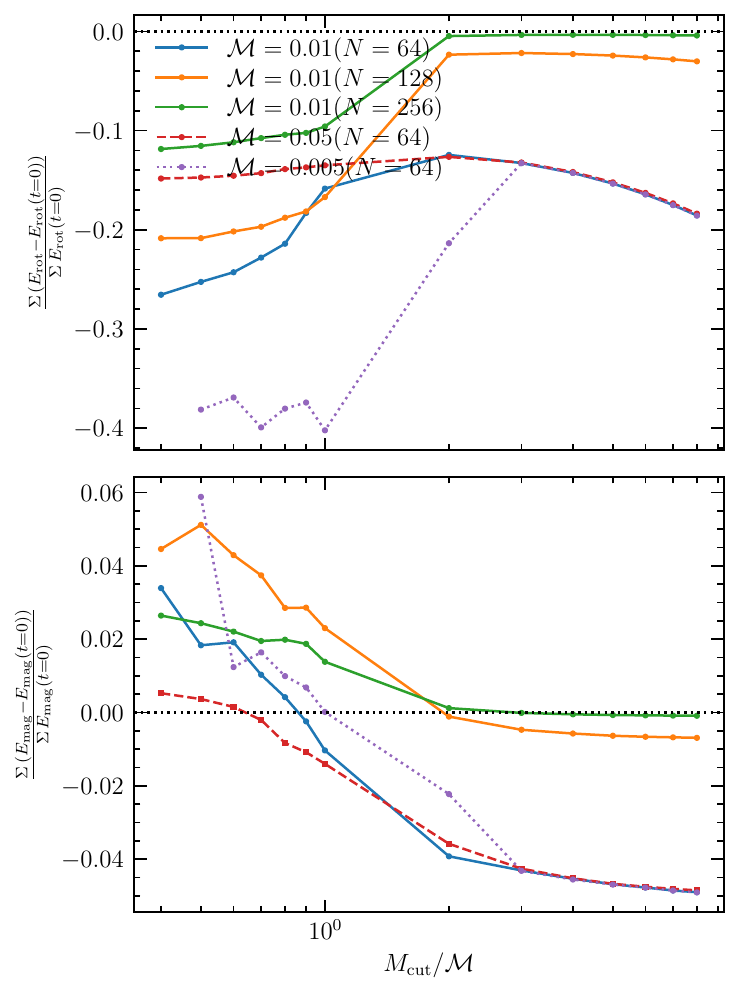}
    \caption{Variation of the amount of rotational (top panel) and magnetic energy (bottom panel) dissipated at the end of one advection as a function of $M_\mathrm{cut}/\mathcal{M}$ at various Mach numbers (M=0.005, 0.01, 0.05) and resolutions (M=0.01 at N=64, 128, 256). For $M_\mathrm{cut}/\mathcal{M}\le1$, there is a spurious increase in the magnetic energy and the dissipation of rotational energy. $M_\mathrm{cut}/\mathcal{M}\gtrsim1$ provides a fairly universal choice that provides stable, physical  solutions with low dissipation. We chose $M_\mathrm{cut}/\mathcal{M}=2$ for our $64\times64$ test at Mach 0.01.}
    \label{fig:mcut_dissipation}
\end{figure}

\begin{figure*}
    \centering
    \includegraphics[width=1\linewidth]{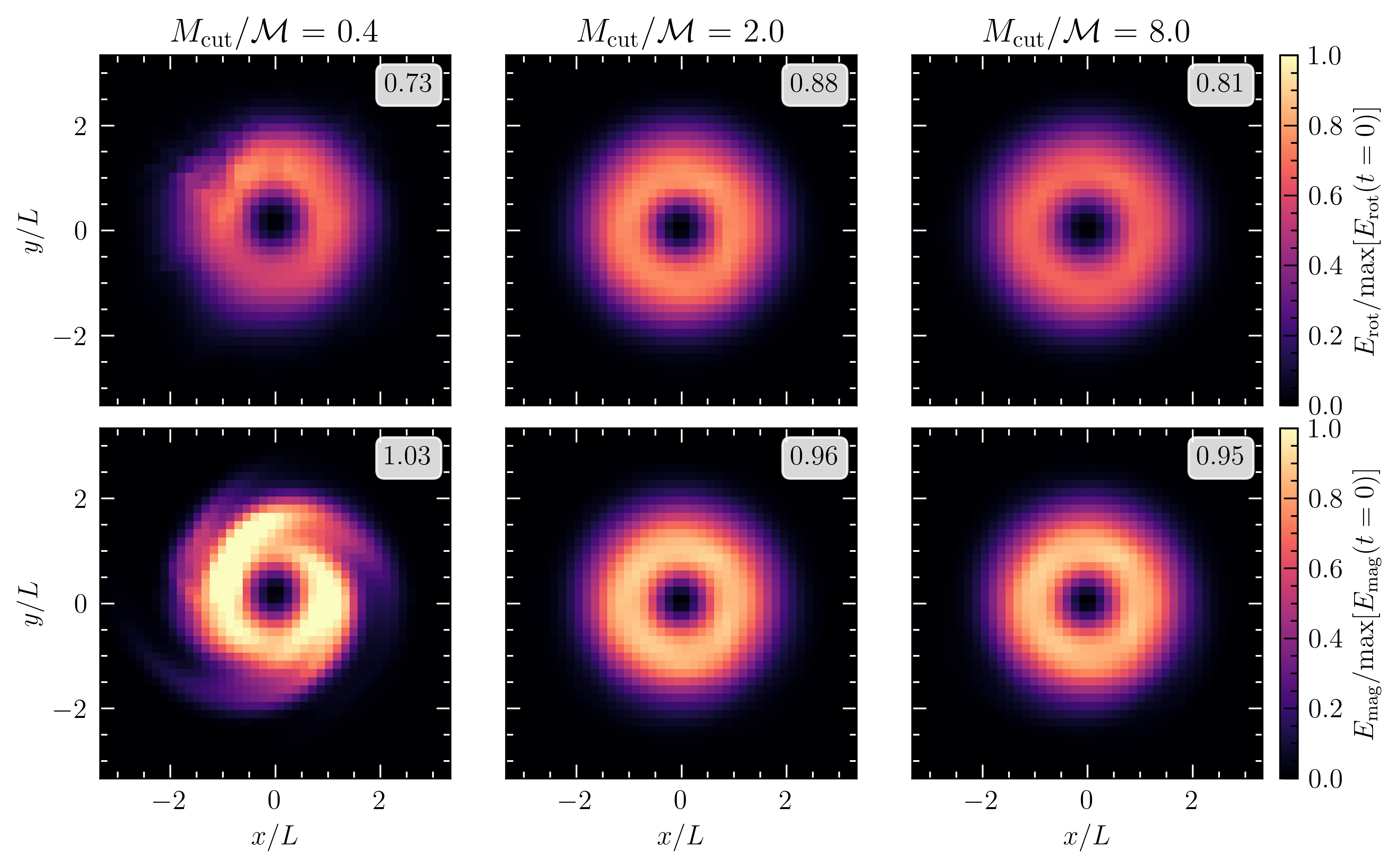}
    \caption{Rotational (top panel) and magnetic energy (bottom panel) profile of the vortex at the end of one advection for $M_\mathrm{cut}/\mathcal{M}=0.4,2~\text{and}~8$. $M_\mathrm{cut}/\mathcal{M}\le1$ shows a spurious increase in the magnetic energy and distorts the vortex. In contrast, $M_\mathrm{cut}/\mathcal{M}\gtrsim1$ yields stable, low-dissipation solutions.}
    \label{fig:mcut_snapshots}
\end{figure*}

Here, we explore the effect of $M_\mathrm{cut}$. We run simulations of the Balsara Vortex test problem at $\mathcal{M}=0.01$ at resolutions of $64^2$, $128^2$ and $256^2$ for $M_\mathrm{cut}/\mathcal{M}$ ranging from 0.4 to 8. We also run additional tests, at $\mathcal{M}=0.005$ and $\mathcal{M}=0.05$ at $64^2$ resolution. Figure~\ref{fig:mcut_dissipation} shows the amount of rotational (top panel) and magnetic energy (bottom panel) dissipated at the end of one complete advection of the vortex. For $M_\mathrm{cut}/\mathcal{M}\le1$, there is an increase in the rotational energy dissipation and a spurious increase in the magnetic energy is also seen. This spurious increase in the magnetic energy corresponds to a significant distortion of the vortex. An example of such a distortion is shown in the left most panel of Figure~\ref{fig:mcut_snapshots}, which shows the profile of the rotational (top panel) and the magnetic energy (bottom panel) corresponding to $M_\mathrm{cut}/\mathcal{M}=0.4, 2~\text{and}~8$ for the 64$\,\times\,$64 run at Mach 0.01.

\begin{figure}
    \centering
    \includegraphics[width=1\linewidth]{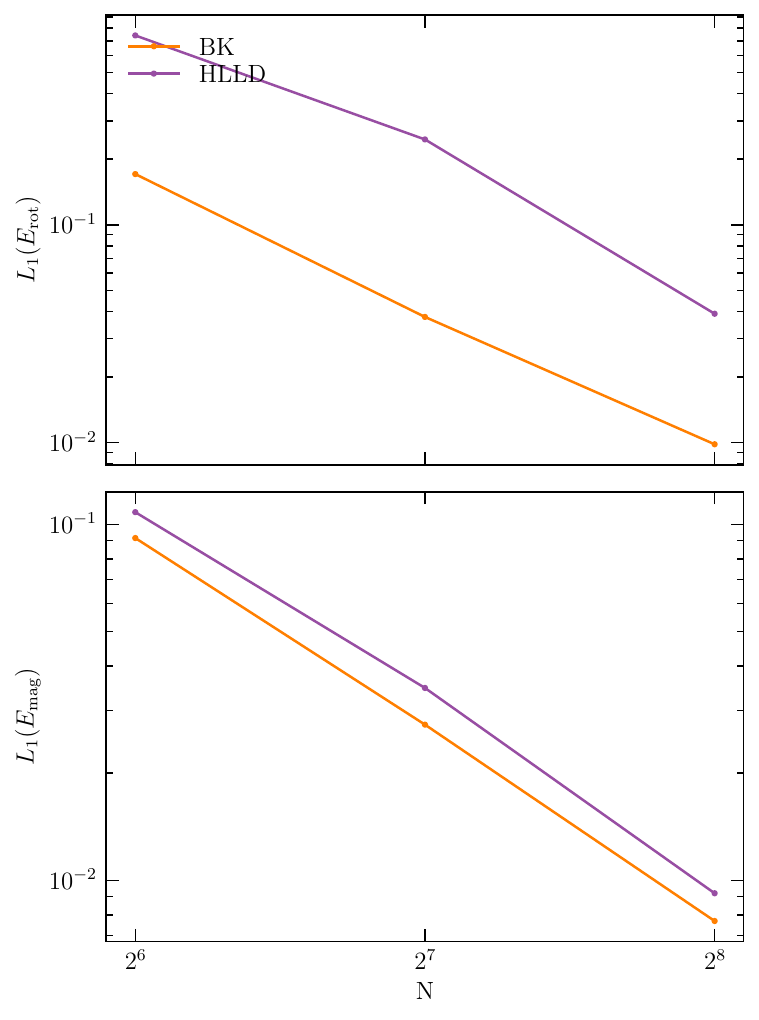}
    \caption{Variation of the $L_1$ norm with increasing resolution. $N$ is the number of resolution elements in the $x/y$ direction and $L_\mathrm{E}$ is calculated via Eq.~\eqref{eq:L1_norm}. USM-BK performs better than USM-HLLD and the error decreases with increasing resolution for both of them, implying that neither show any growth of dispersive errors, with $M_\mathrm{cut}/\mathcal{M}=2$ is a good choice for USM-BK.}
    \label{fig:l1_res}
\end{figure}

In Figure~\ref{fig:mcut_dissipation}, $M_\mathrm{cut}/\mathcal{M}=2$ shows the least dissipation of rotational kinetic energy and conserves the magnetic energy well. Therefore, we choose this value for our vortex test. We also perform a resolution test measuring the $L_\mathrm{1}$ norm of the rotational kinetic energy and the magnetic energy for USM-BK (at $M_\mathrm{cut}/\mathcal{M}=2$) and USM-HLLD. The $L_\mathrm{1}$ for energy $E$ (rotational or magnetic) is calculated using
\begin{equation}
    L_\mathrm{1}(E) = \frac{\Sigma|E-E(t=0)|}{\Sigma E(t=0)}\,.
    \label{eq:L1_norm}
\end{equation}
The variation of the $L_\mathrm{1}$ norm with the number of resolution elements in the $x/y$ direction ($N$) are shown in Figure~\ref{fig:l1_res}. USM-HLLD has a higher value of $L_\mathrm{1}$ norm compared to USM-BK and it decreases with increasing resolution for both of them. This shows that USM-BK performs better than USM-HLLD and there is no growth of dispersive errors.

It is clear that the optimal choice of $M_\mathrm{cut}/\mathcal{M}$ is Mach number-dependent and may depend on the type of problem too. For a given MHD problem, the optimal value can be inferred from low-resolution runs before the solver is used in the corresponding high-resolution production run. Nevertheless, the parameter variation tests in Fig.~\ref{fig:mcut_dissipation} suggest that $M_\mathrm{cut}/\mathcal{M}\gtrsim1$ provides a relatively universal choice that yields stable and low-dissipation solutions with USM-BK.

\section{Effective hydrodynamic and magnetic Reynolds numbers}
\label{appendix:Pm k_eta rel}

The hydrodynamic Reynolds number ($\mathrm{Re}$) is defined as
\begin{equation}
    \mathrm{Re}=\frac{u_\mathrm{turb}\ell_\mathrm{turb}}{\nu},
\end{equation}
where $u_\mathrm{turb}$ is the fluid turbulent velocity at the driving scale of turbulence ($\ell_\mathrm{turb}=2\pi/k_\mathrm{driving}$) and $\nu$ is the kinematic viscosity of the fluid.
Similarly, the magnetic Reynolds number ($\mathrm{Rm}$) is defined as
\begin{equation}
    \mathrm{Rm}=\frac{u_\mathrm{turb}\ell_\mathrm{turb}}{\eta},
\end{equation}
where $\eta$ is the magnetic resistivity of the fluid.

Numerical viscosity and resistivity must be lower than the explicit viscosity and resistivity to avoid smearing of features and over-damping of flows. Therefore, the numerical Reynolds numbers associated with a numerical scheme must be greater than the explicit Reynolds number being simulated. We can calculate the numerical Hydrodynamic Reynolds number and Magnetic Prandtl number of ideal MHD simulations from measurements of the characteristic dissipation scales.

To calculate the hydrodynamic Reynolds number, we use the following relation given in \citet{KrielEtAl2022}
\begin{equation}
    \mathrm{Re}=\left(\frac{k_\nu}{c_\nu k_\mathrm{driving}}\right)^{4/3},
    \label{eq:k_nu}
\end{equation}
where $c_\nu=0.025^{+0.005}_{-0.006}$ (referred to as $c_\mathrm{Re}$ in the main text).

To measure the magnetic Reynolds number, we follow \citet{KrielEtAl2023}. However, their definition of the resistive dissipation scale (their definition will be referred to as $k_\nu'$) is different from what we have used in Eq.~(\ref{eq:Pkin}). They define the viscous dissipation wave-number as the wave-number where the scale-dependent hydrodynamic Reynolds number equals one, i.e., $\mathrm{Re}(k_\nu')=1$. This wave-number marks the scale where the flow transitions from an inertial force dominated one ($k_\mathrm{turb}<k<k_\nu'$) to a dissipation-dominated one ($k>k_\nu'$). \citet{KrielEtAl2023} have shown that $k_\nu'$ and $\mathrm{Re}$ are related by
\begin{equation}
    k_\nu'=c_\nu'k_\mathrm{driving}\mathrm{Re}^{3/4},
    \label{eq:k_nu prime}
\end{equation}
where $c_\nu'=0.10^{+0.01}_{-0.01}$.

Using Eq.~(\ref{eq:k_nu}) and Eq.~(\ref{eq:k_nu prime}), we can write
\begin{equation}
    k_\nu'=\frac{c_\nu'}{c_\nu}k_\nu.
    \label{eq:k_nu k_nu prime relation}
\end{equation}
Thus, if we have measured $k_\nu$ from spectral fitting in Eq.~(\ref{eq:Pkin}), we can find $k_\nu'$.

\citet{KrielEtAl2023} have also shown the resistive dissipation wave-number ($k_\eta$) scales with the viscous dissipation wave-number ($k_\nu'$, based on the alternative definition mentioned here) and the Prandtl number as
\begin{equation}
    k_\eta=c_\eta k_\nu'\mathrm{Pm}^{1/2},
    \label{eq:k_eta k_nu prime relation}
\end{equation}
where $c_\eta=0.53^{+0.07}_{-0.07}$.

Using Eq.~(\ref{eq:k_nu k_nu prime relation}) and Eq.~(\ref{eq:k_eta k_nu prime relation}), we can write
\begin{equation}
    k_\eta=c_\eta\frac{c_\nu'}{c_\nu}k_\nu\mathrm{Pm}^{1/2}=c_\mathrm{Pm}k_\nu\mathrm{Pm}^{1/2},
\end{equation}
where $c_\mathrm{Pm}=c_\eta c_\nu'/c_\nu=2.1^{+0.8}_{-0.5}$.

The Prandtl number (Pm) can be obtained using the above relation. The Magnetic Reynolds number can then be calculated as
\begin{equation}
    \mathrm{Rm} = \mathrm{Re} \times \mathrm{Pm}.
\end{equation}
%%%%%%%%%%%%%%%%%%%%%%%%%%%%%%%%%%%%%%%%%%%%%%%%%%

% Don't change these lines
\bsp	% typesetting comment
\label{lastpage}
\end{document}